\documentclass[12pt, draftclsnofoot, onecolumn]{IEEEtran}

\ifCLASSINFOpdf
  \usepackage[pdftex]{graphicx}
  \DeclareGraphicsExtensions{.pdf,.jpeg,.png}
\else
  \usepackage[dvips]{graphicx}
\fi
\usepackage{epstopdf}
\usepackage{amssymb,amsmath,amsthm}
\usepackage{amssymb}
\usepackage{wasysym}
\usepackage{algorithm}
\usepackage{algorithmic}
\usepackage{array}
\usepackage{cite}
\usepackage{color}
\usepackage{url}

\usepackage{epsfig,latexsym}
\usepackage{flushend}
\usepackage{cite}
\usepackage{verbatim}
\usepackage{amsopn}
\usepackage{booktabs}

\usepackage{stfloats}
\usepackage{amsmath}

\usepackage{subfigure}
\usepackage{hyperref}
\usepackage{mathrsfs}
\usepackage{setspace}

\begin{document}

\title{MIMO Capacity Characterization for Movable Antenna Systems}
\author{{Wenyan Ma, \IEEEmembership{Student Member, IEEE}, Lipeng Zhu, \IEEEmembership{Member, IEEE}, and  Rui Zhang, \IEEEmembership{Fellow, IEEE}}
	\vspace{-25pt}
	
	\thanks{W. Ma and L. Zhu are with the Department of Electrical and Computer Engineering, National University of Singapore,
		Singapore 117583 (Email: {wenyan@u.nus.edu}, {zhulp@nus.edu.sg}).
		
		R. Zhang is with The Chinese University of Hong Kong, Shenzhen, and Shenzhen Research Institute of Big Data, Shenzhen,
		China 518172 (Email: rzhang@cuhk.edu.cn). He is also with the Department of Electrical and Computer Engineering, National
		University of Singapore, Singapore 117583 (Email: elezhang@nus.edu.sg).
}}
\maketitle

\begin{abstract}
	In this paper, we propose a new multiple-input multiple-output (MIMO) communication system with movable antennas
	(MAs) to exploit the antenna position optimization for enhancing the capacity. Different from conventional MIMO systems with fixed-position antennas (FPAs), the proposed system can flexibly change the positions of transmit/receive MAs, such that the MIMO channel between them is reconfigured to achieve higher capacity. We aim to characterize the capacity of MA-enabled point-to-point MIMO communication systems, by jointly optimizing the positions of transmit and receive MAs as well as the covariance of transmit signals. First, we develop an efficient alternating optimization algorithm to find a locally optimal solution by iteratively optimizing the transmit covariance matrix and the position of each transmit/receive MA   with the other variables being fixed. Next, we propose alternative algorithms of lower complexity for capacity maximization in the low-SNR regime and for the multiple-input single-output (MISO) and single-input multiple-output (SIMO) cases. Numerical results show that our proposed MA systems significantly improve the MIMO channel capacity compared to traditional FPA systems as well as various benchmark schemes, and useful insights are drawn into the capacity gains of MA systems.

\end{abstract}

\section{Introduction}

Driven by the explosive growth of wireless applications, such as high-quality video streaming and virtual/augmented reality (VR/AR), there is an ever-increasing demand for higher capacity in the future sixth-generation (6G) and beyond wireless communication systems. To achieve this goal, multiple-input multiple-output (MIMO) and massive MIMO communications have been the key enabling technologies \cite{jiang2021the,saad2020a}. However, although MIMO/massive MIMO can enhance the wireless channel capacity and system spectral efficiency significantly, they generally require considerably higher energy consumption and hardware cost, due to the increasingly more antennas and/or radio frequency (RF) chains required for wireless systems operating at higher frequency bands \cite{chowdhury20206g,zhu2019milli}.

To reduce the number of RF chains, antenna selection (AS) is a practical solution for capturing a large portion of the channel
capacity in MIMO systems by properly selecting a small number of antennas with favorable channels from the a large candidate set \cite{lu2014an,sanayei2004antenna}. However, deploying more candidate antennas requires higher hardware cost, and the overhead of channel estimation and the computational complexity of AS algorithms grow greatly with the total number of antennas \cite{gharavi2004fast}. Moreover, since the antennas are deployed at fixed positions in conventional MIMO/massive MIMO with/without AS, they cannot fully utilize the spatial variation of wireless channels in a given transmit/receive region, especially with a limited number of antennas.

To further explore the degrees of freedom (DoFs) in the spatial domain for improving the communication performance, the fluid antenna system (FAS) was proposed by changing the antenna position flexibly over a one-dimensional (1D) line \cite{wong2021fluid,wong2022fluid}. By using conductive fluids as materials for antennas, the position of a receive antenna can be switched freely among all candidate ports over a fixed-length line, and thus the signal with the highest signal-to-noise ratio (SNR) can be received \cite{wong2021fluid}. By setting the number of possible positions large enough, a single-antenna FAS can even achieve lower outage probability compared to multiple-antenna systems with maximum ratio combining (MRC). Moreover, the fluid antenna multiple access (FAMA) was studied in \cite{wong2022fluid} to support multiple transceivers with a single fluid antenna at each mobile user. By selecting the positions of the fluid antennas at different users, the favorable channel condition with mitigated interference can be obtained, thereby achieving a significant capacity gain. It was shown in \cite{wong2021fluid,wong2022fluid} that changing the positions of antennas can efficiently improve the wireless channel capacity. However, due to the limitations of liquid materials, the FAS can only support a single fluid antenna moving over a 1D line, which limits its capability for fully exploiting the wireless channel variation in the spatial domain.

On the other hand, to fully explore the ultimate capacity of MIMO systems, the concept of continuous-aperture MIMO (CAP-MIMO) was proposed \cite{zhang2022pattern}, which is also called holographic MIMO \cite{huang2020holo,pizzo2022fourier}, large intelligent surface \cite{decarli2021access}, or holographic surface \cite{deng2021reconfi,wei2022multi}. By deploying a large number of sub-wavelength radiation elements in a compact surface, CAP-MIMO can be regarded as a quasi-continuous electromagnetic surface with controllable current density. Although the recent
advance in highly-flexible programmable meta-materials has made the CAP-MIMO more practically implementable, some critical challenges still remain unsolved \cite{huang2020holo}, such as the efficient estimation of ultra high-dimensional CAP-MIMO channels and the efficient current density optimization subject to a large number of practical constraints.

To exploit more spatial DoFs for enhancing the channel capacity, we propose a new MIMO communication system enabled by movable antennas (MAs). Specifically, by connecting the MAs to RF chains via flexible cables, the positions of MAs can be adjusted by controllers in real time, such as stepper motors or servos \cite{ismial1991null,basbug2017design,zhuravlev2015experi}. Different from conventional MIMO systems with fixed-position antennas (FPAs), the proposed system can flexibly change the positions of transmit/receive MAs, such that the MIMO channel matrix between them can be reshaped for achieving higher capacity. For MIMO systems, we can jointly optimize the positions of transmit and receive MAs for the channel capacity maximization. Since the MIMO channel capacity is generally achieved by transmitting multiple data streams in parallel, the positions of transmit and receive MAs need to be designed to optimally balance the channel gains for multiple spatial data streams so as to maximize the channel capacity. 

In this paper, we study the joint optimization of the transmit and receive MAs' positions as well as the transmit signals' covariance matrix for maximizing
the capacity of a point-to-point MA-enabled MIMO system, with MAs equipped at both the transmitter and receiver. To characterize the fundamental capacity limit, we assume that perfect channel state information (CSI) is available at both the transmitter and receiver. Our main contributions are summarized as follows:

\begin{itemize}
	\item First, we investigate the capacity maximization problem for a MIMO communication system enabled by MAs, which is non-convex and thus difficult to be optimally solved. To tackle this problem efficiently, we propose an alternating optimization algorithm by iteratively optimizing the transmit covariance matrix and the position of each transmit/receive MAs with the other variables being fixed. In particular, given the transmit covariance matrix, we leverage the convex relaxation technique to find a suboptimal solution for the transmit/receive MA-positioning optimization subproblem. It is shown that the proposed algorithm converges to a locally optimal solution at least.
	\item Moreover, we derive the MA-enabled MIMO channel capacity in the asymptotically low-SNR regime, and propose an alternative algorithm for solving the capacity maximization problem with lower complexity. In addition, we further simplify the proposed algorithm for the capacity maximization in the special cases of multiple-input single-output (MISO) and single-input multiple-output (SIMO) channels with MAs.
	\item Finally, we provide extensive numerical results to demonstrate the capacity gains of our proposed MA-enabled MIMO systems over conventional MIMO systems with FPAs, and shed more light on the performance gains of MA systems. In particular, it is shown that by jointly designing the positions of transmit and receive MAs, the total MIMO channel power can be significantly improved while the condition number of the MIMO channel decreases with the size of the transmit/receive region, thus leading to higher MIMO channel capacity.
\end{itemize}

%\begin{table}[!b]
%	\centering
%	\caption{Symbol notations.}
%	\label{Talbe1}
%	\begin{tabular}{|p{1.4cm}|p{5.7cm}|p{1.4cm}|p{5.7cm}|}
%		\hline
%		\textbf{Symbol}  &   \textbf{Description}     &   \textbf{Symbol}      &   \textbf{Description}         \\
%		\hline
%		$N$  &   Number of transmit MAs  &   $M$  &   Number of receive MAs  \\
%		\hline
%		 $\{\boldsymbol{t}_n\}_{n=1}^N$  &    Positions of transmit MAs & $\{\boldsymbol{r}_m\}_{m=1}^M$  &   Positions of receive MAs   \\\hline
%		$\boldsymbol{Q}$  &  Transmit covariance matrix   &  $\boldsymbol{H}(\tilde{\boldsymbol{t}}, \tilde{\boldsymbol{r}})$  & Channel matrix  \\
%		\hline
%		$\boldsymbol{G}(\tilde{\boldsymbol{t}})$  &  Field response matrix of transmit region   &   $\boldsymbol{F}(\tilde{\boldsymbol{r}})$  & Field response matrix of receive region  \\
%		\hline
%		$\boldsymbol{\Sigma}$  &  Path response matrix  &  $D$  &   Minimum distance between antennas  \\
%		\hline
%	\end{tabular}
%\end{table}

\begin{table}[!b]
	\centering
	\caption{Symbol notations.}
	\label{Talbe1}
	\begin{tabular}{|p{1.4cm}|p{5.7cm}|p{1.4cm}|p{5.7cm}|}
		\hline
		\textbf{Symbol}  &   \textbf{Description}     &   \textbf{Symbol}      &   \textbf{Description}         \\
		\hline
		$N$  &   Number of transmit MAs  &   $M$  &   Number of receive MAs  \\
		\hline
		$\mathcal{C}_t$ &  Transmit region  &   $\mathcal{C}_r$  &   Receive region   \\
		\hline
		$L_t$  &   Number of transmit paths  &  $L_r$  &   Number of receive paths   \\
		\hline
		$\{\boldsymbol{t}_n\}_{n=1}^N$  &    Positions of transmit MAs & $\{\boldsymbol{r}_m\}_{m=1}^M$  &   Positions of receive MAs   \\\hline
		$\boldsymbol{Q}$  &  Transmit covariance matrix   &  $\boldsymbol{H}(\tilde{\boldsymbol{t}}, \tilde{\boldsymbol{r}})$  & Channel matrix  \\
		\hline
		$\boldsymbol{G}(\tilde{\boldsymbol{t}})$  &  Field response matrix of transmit region   &   $\boldsymbol{F}(\tilde{\boldsymbol{r}})$  & Field response matrix of receive region  \\
		\hline
		$\{\theta_t^p\}_{p=1}^{L_t}$  &   Elevation AoDs of $L_t$ transmit paths  &  $\{\phi_t^p\}_{p=1}^{L_t}$  &   Azimuth AoDs of $L_t$ transmit paths   \\
		\hline
		$\{\theta_r^q\}_{q=1}^{L_r}$  &   Elevation AoAs of $L_r$ receive paths  &  $\{\phi_r^q\}_{q=1}^{L_r}$  &   Azimuth AoAs of $L_r$ receive paths   \\
		\hline
		$P$  &  Transmit power  &  $\sigma^2$  &   Average noise power   \\
		\hline
		$\boldsymbol{\Sigma}$  &  Path response matrix  &  $D$  &   Minimum distance between antennas  \\
		\hline
	\end{tabular}
\end{table}

The rest of this paper is organized as follows. Section II presents the system model and the problem formulation for characterizing the MA-enabled MIMO system capacity. Section III provides an alternating optimization algorithm for solving the formulated problem under different setups. Numerical results and discussions are presented in Section IV. Finally, Section V concludes this paper.

\textit{Notations}: Symbols for vectors (lower case) and matrices (upper case) are in boldface.  $(\cdot)^T $ and $(\cdot)^H $ denote the transpose and conjugate transpose (Hermitian), respectively. The set of $P\times{Q}$ dimensional complex and real matrices is denoted by $\mathbb{C}^{P\times{Q}}$ and $\mathbb{R}^{P\times{Q}}$, respectively. We use $\boldsymbol{a}[p]$ and $\boldsymbol{A}[p,q]$ to denote the $p$th entry of vector $\boldsymbol{a}$ and the entry of matrix $\boldsymbol{A}$ in its $p$th row and $q$th column, respectively. $\rm{Re}\{\boldsymbol{a}\}$ denotes the real part of vector $\boldsymbol{a}$. The trace of matrix $\boldsymbol{A}$ is denoted by $\rm {Tr}(\boldsymbol{A})$. $\textrm{diag}(\boldsymbol{a})$ denotes a square diagonal matrix with the elements of vector $\boldsymbol{a}$ on the main diagonal, while $\textrm{diag}(\boldsymbol{A})$ denotes a column vector consisting of the main diagonal elements of matrix $\boldsymbol{A}$. $\boldsymbol{A} \succeq 0$ indicates that $\boldsymbol{A}$ is a positive semi-definite matrix. $\mathcal{CN}(0,\boldsymbol{\Gamma})$ denotes the circularly symmetric complex Gaussian (CSCG) distribution with mean zero and covariance matrix $\boldsymbol{\Gamma}$. We use $\boldsymbol{I}_{K}$ to represent the identity matrix of order $K$. The $2$-norm of vector $\boldsymbol{a}$ is denoted by $\|\boldsymbol{a}\|_2$. $\|\boldsymbol{A}\|_2$ and $\|\boldsymbol{A}\|_F$ denote the spectral norm and Frobenius norm of matrix $\boldsymbol{A}$, respectively. $\textrm{rank}(\boldsymbol{A})$ denotes the rank of matrix $\boldsymbol{A}$. The amplitude and phase of complex number $a$ are denoted by $|a|$ and $\angle{a}$, respectively. For ease of reference, the main symbols used in this paper are listed in Table \ref{Talbe1}.

\section{System Model and Problem Formulation}
\subsection{MA-Enabled MIMO System }

\begin{figure}[!t]
	\centering
	\includegraphics[width=160mm]{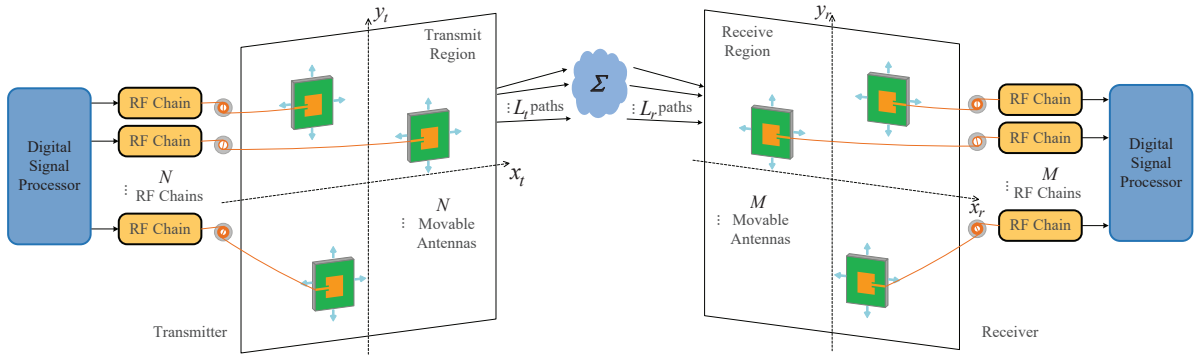}
	\caption{The MA-enabled MIMO communication system.}
	\label{FIG1}
\end{figure}

We consider a MIMO communication system with $N$ MAs at the transmitter and $M$ MAs at the receiver, as shown in Fig.~\ref{FIG1}. The transmit and receive MAs are connected to RF chains via flexible cables, and thus their positions can be adjusted in real time. The positions of the $n$th ($n=1,2,\ldots,N$) transmit MA and $m$th ($m=1,2,\ldots,M$) receive MA can be represented by Cartesian coordinates $\boldsymbol{t}_n=[x_{t,n}, y_{t,n}]^T \in \mathcal{C}_t$ and $\boldsymbol{r}_m=[x_{r,m}, y_{r,m}]^T \in \mathcal{C}_r$, where $\mathcal{C}_t$ and $\mathcal{C}_r$ denote the given two-dimensional (2D) regions within which the transmit and receive MAs can move freely, respectively. Without loss of generality, we set $\mathcal{C}_t$ and $\mathcal{C}_r$ as square regions with size $A \times A$. 

We assume quasi-static block-fading channels, and focus on one particular fading block with the multi-path channel components at any location in $\mathcal{C}_t$/$\mathcal{C}_r$ given as fixed. For MA-enabled MIMO communication systems, the channel is reconfigurable by adjusting the positions of transmit and receive MAs. Denote the collections of the coordinates of $N$ transmit MAs and $M$ receive MAs by $\tilde{\boldsymbol{t}}=\left[\boldsymbol{t}_1, \boldsymbol{t}_2, \ldots, \boldsymbol{t}_N\right] \in{\mathbb{R}^{2\times{N}}}$ and $\tilde{\boldsymbol{r}}=\left[\boldsymbol{r}_1, \boldsymbol{r}_2, \ldots, \boldsymbol{r}_M\right] \in{\mathbb{R}^{2\times{M}}}$, respectively. Then, the MIMO channel matrix from the transmitter to the receiver is given by $\boldsymbol{H}(\tilde{\boldsymbol{t}}, \tilde{\boldsymbol{r}}) \in{\mathbb{C}^{M\times{N}}}$, which is a function of $\tilde{\boldsymbol{t}}$ and $\tilde{\boldsymbol{r}}$ in general.

Let $\boldsymbol{s}\in \mathbb{C}^{N}$ denote the transmit signal vector. The corresponding covariance matrix is defined as $\boldsymbol{Q} \triangleq \mathbb{E}\{\boldsymbol{s} \boldsymbol{s}^H\} \in \mathbb{C}^{N \times N}$, with $\boldsymbol{Q} \succeq \mathbf{0}$. We consider an average sum power constraint at the transmitter given by $\mathbb{E}\{\|\boldsymbol{s}\|_2^2\} \leq P$, which is equivalent to ${\rm {Tr}(\boldsymbol{Q})} \leq P$. The received signal vector is thus given by
\begin{equation}\label{y}
  \boldsymbol{y}(\tilde{\boldsymbol{t}}, \tilde{\boldsymbol{r}}) = \boldsymbol{H}(\tilde{\boldsymbol{t}}, \tilde{\boldsymbol{r}}) \boldsymbol{s} + \boldsymbol{z},
\end{equation}
where $\boldsymbol{z} \sim \mathcal{CN}(0,\sigma^2\boldsymbol{I}_M)$ denotes the additive white Gaussian noise (AWGN) vector at the receiver, which is assumed to be CSCG distributed with zero mean, and $\sigma^2$ being the average noise power.

\subsection{Field-Response Based Channel Model}

\begin{figure}[!t]
\centering
\includegraphics[width=110mm]{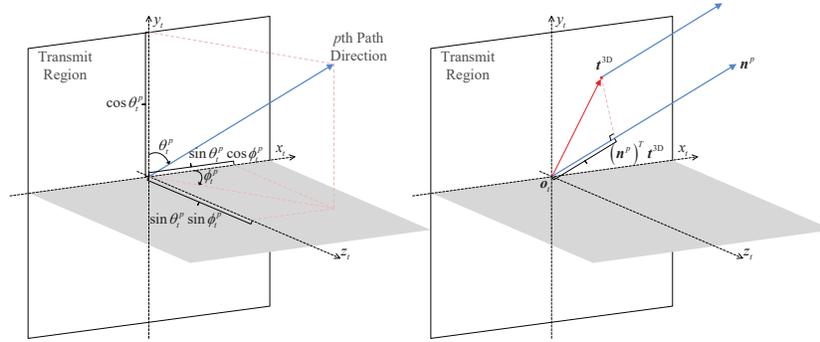}
\caption{The spatial angles and path differences for transmit region.}
\label{FIG2}
\end{figure}

For the MA-enabled MIMO communication system, the channel matrix is determined by the signal propagation environment and the positions of transmit and receive MAs. We consider the far-field wireless channel model, where the size of the transmit/receive region is much smaller than the signal propagation distance. Thus, for each channel path component, all MAs in the transmit/receive region experience the same angle of departure (AoD)/angle of arrival (AoA), and amplitude of the complex path coefficient, while the phase of the complex path coefficient varies for different transmit/receive MAs at different positions.

Denote the number of transmit paths and receive paths as $L_t$ and $L_r$, respectively. As shown in Fig.~\ref{FIG2}, the elevation and azimuth AoDs of the $p$th ($p=1,2,\ldots,L_t$) transmit path are denoted by $\theta_t^p \in [0, \pi]$ and $\phi_t^p \in [0, \pi]$, respectively. Then, the normalized wave vector of the $p$th transmit path can be represented by $\boldsymbol{n}^p = [\sin \theta_t^p \cos \phi_t^p, \cos \theta_t^p, \sin \theta_t^p \sin \phi_t^p]^T$. Denote the three-dimensional (3D) coordinate vector of the transmit MA and origin of the transmit region by $\boldsymbol{t}^{\rm 3D}=[x_{t}, y_{t}, 0]^T$ and $\boldsymbol{o}_t = [0, 0, 0]^T$, respectively. Then, the difference of the signal propagation for the $p$th transmit path between position $\boldsymbol{t}^{\rm 3D}$ and $\boldsymbol{o}_t$ is given by
\begin{equation}\label{rho}
  \rho_t^p(\boldsymbol{t}) = (\boldsymbol{n}^p)^T(\boldsymbol{t}^{\rm 3D}-\boldsymbol{o}_t) = x_{t} \sin \theta_t^p \cos \phi_t^p + y_{t} \cos \theta_t^p.
\end{equation}
Thus, the phase difference of the $p$th transmit path between $\boldsymbol{t}^{\rm 3D}$ and $\boldsymbol{o}_t$ is $2\pi\rho_t^p(\boldsymbol{t})/\lambda$, where $\lambda$ is the carrier wavelength. The field response vector of the transmit MA can be defined as
\begin{equation}\label{g}
  \boldsymbol{g}(\boldsymbol{t}) \triangleq \left[ e^{j\frac{2\pi}{\lambda}\rho_t^1(\boldsymbol{t})}, e^{j\frac{2\pi}{\lambda}\rho_t^2(\boldsymbol{t})}, \ldots, e^{j\frac{2\pi}{\lambda}\rho_t^{L_t}(\boldsymbol{t})} \right]^T \in{\mathbb{C}^{L_t}}.
\end{equation}
By stacking $\boldsymbol{g}(\boldsymbol{t}_n)$ of all $N$ transmit MAs, the field response matrix of all the $N$ transmit MAs is given by
\begin{equation}\label{G}
  \boldsymbol{G}(\tilde{\boldsymbol{t}})\triangleq\left[\boldsymbol{g}\left(\boldsymbol{t}_1\right), \boldsymbol{g}\left(\boldsymbol{t}_2\right), \ldots, \boldsymbol{g}\left(\boldsymbol{t}_N\right)\right] \in \mathbb{C}^{L_t \times N}.
\end{equation}

Similarly, denote the elevation and azimuth AoAs of the $q$th ($q=1,2,\ldots,L_r$) receive path as $\theta_r^q \in [0, \pi]$ and $\phi_r^q \in [0, \pi]$, respectively. The field response vector of the receive MA is defined as
\begin{equation}\label{f}
  \boldsymbol{f}(\boldsymbol{r}) \triangleq \left[ e^{j\frac{2\pi}{\lambda}\rho_r^1(\boldsymbol{r})}, e^{j\frac{2\pi}{\lambda}\rho_r^2(\boldsymbol{r})}, \ldots, e^{j\frac{2\pi}{\lambda}\rho_r^{L_r}(\boldsymbol{r})} \right]^T \in{\mathbb{C}^{L_r}},
\end{equation}
where $\rho_r^q(\boldsymbol{r}) = x_{r} \sin \theta_r^q \cos \phi_r^q + y_{r} \cos \theta_r^q$ is the difference of the signal propagation distance for the $q$th receive path between position $\boldsymbol{r}$ and the origin of the receive region. Thus, the field response matrix of all the $M$ receive MAs is written as
\begin{equation}\label{F}
  \boldsymbol{F}(\tilde{\boldsymbol{r}})\triangleq\left[\boldsymbol{f}\left(\boldsymbol{r}_1\right), \boldsymbol{f}\left(\boldsymbol{r}_2\right), \ldots, \boldsymbol{f}\left(\boldsymbol{r}_M\right)\right] \in \mathbb{C}^{L_r \times M}.
\end{equation}

Furthermore, we define the path response matrix from the origin of the transmit region to that of the receive region as $\boldsymbol{\Sigma} \in{\mathbb{C}^{L_r \times L_t}}$, where $\boldsymbol{\Sigma}[q,p]$ is the response between the $p$th transmit path and the $q$th receive path. As a result, the channel matrix from the transmitter to the receiver is given by
\begin{equation}\label{H}
  \boldsymbol{H}(\tilde{\boldsymbol{t}}, \tilde{\boldsymbol{r}}) = \boldsymbol{F}(\tilde{\boldsymbol{r}})^H \boldsymbol{\Sigma} \boldsymbol{G}(\tilde{\boldsymbol{t}}).
\end{equation}

\subsection{Problem Formulation}
To reveal the fundamental capacity limit of the MA-enabled MIMO communication system, we assume that perfect CSI is available at both the transmitter and receiver. The MIMO channel capacity is thus given by
\begin{align}\label{C}
  C(\tilde{\boldsymbol{t}}, \tilde{\boldsymbol{r}})=\max_{\boldsymbol{Q}: {\rm {Tr}}(\boldsymbol{Q}) \leq P, \boldsymbol{Q} \succeq \mathbf{0}} \log_{2} \det\left(\boldsymbol{I}_M+\frac{1}{\sigma^2} \boldsymbol{H}(\tilde{\boldsymbol{t}}, \tilde{\boldsymbol{r}}) \boldsymbol{Q} \boldsymbol{H}(\tilde{\boldsymbol{t}}, \tilde{\boldsymbol{r}})^H\right)
\end{align}
in bits per second per Hertz (bps/Hz). It is worth noting that different from the conventional MIMO channel with FPAs, the capacity for the MA-enabled MIMO channel shown in \eqref{C} is depended on the positions of transmit and receive MAs $\tilde{\boldsymbol{t}}$ and $\tilde{\boldsymbol{r}}$, which influence the channel matrix $\boldsymbol{H}(\tilde{\boldsymbol{t}}, \tilde{\boldsymbol{r}})$ as well as the corresponding optimal transmit covariance matrix $\boldsymbol{Q}$.

In order to avoid the coupling effect between the antennas in the transmit/receive region, a minimum distance $D$ is required between each pair of antennas, i.e., $\|\boldsymbol{t}_k-\boldsymbol{t}_l\|_2 \geq D$, $k,l = 1,2,\ldots,N$, $k\neq l$, and $\|\boldsymbol{r}_k-\boldsymbol{r}_l\|_2 \geq D$, $k,l = 1,2,\ldots,M$, $k\neq l$. Then, we aim to maximize the capacity of an MA-enabled MIMO channel by jointly optimizing the MA positions $\tilde{\boldsymbol{t}}$, $\tilde{\boldsymbol{r}}$ and the transmit covariance matrix $\boldsymbol{Q}$, subject to the minimum distance constraints on the MA positions and a sum power constraint at the transmitter. The optimization problem is thus formulated as
\begin{subequations}
\begin{align}
\textrm {(P1)}~~\max_{\tilde{\boldsymbol{t}}, \tilde{\boldsymbol{r}}, \boldsymbol{Q}} \quad & \log_{2} \det\left(\boldsymbol{I}_M+\frac{1}{\sigma^2} \boldsymbol{H}(\tilde{\boldsymbol{t}}, \tilde{\boldsymbol{r}}) \boldsymbol{Q} \boldsymbol{H}(\tilde{\boldsymbol{t}}, \tilde{\boldsymbol{r}})^H\right) \label{P1a}\\
\text{s.t.} \quad & \tilde{\boldsymbol{t}} \in \mathcal{C}_t, \label{P1b}\\
                  & \tilde{\boldsymbol{r}} \in \mathcal{C}_r, \label{P1c}\\
                  & \|\boldsymbol{t}_k-\boldsymbol{t}_l\|_2 \geq D,~~ k,l = 1,2,\ldots,N,~~ k\neq l,\label{P1d}\\
                  & \|\boldsymbol{r}_k-\boldsymbol{r}_l\|_2 \geq D,~~ k,l = 1,2,\ldots,M,~~ k\neq l,\label{P1e}\\
& {\rm {Tr}}(\boldsymbol{Q}) \leq P,  \label{P1f}\\
& \boldsymbol{Q} \succeq \mathbf{0}. \label{P1g}
\end{align}
\end{subequations}

Note that problem (P1) is a non-convex optimization problem because the objective function is non-concave over the MA positions $\tilde{\boldsymbol{t}}$ and $\tilde{\boldsymbol{r}}$, and the minimum distance constraints in \eqref{P1d} and \eqref{P1e} are non-convex. Moreover, the transmit covariance matrix $\boldsymbol{Q}$ is coupled with $\tilde{\boldsymbol{t}}$ and $\tilde{\boldsymbol{r}}$ in the objective function of (P1), which makes (P1) challenging to solve.

\section{Proposed Algorithm}
In this section, we propose an alternating optimization algorithm for solving (P1). Specifically, we first transfer the objective function of (P1) into a more tractable form with respect to the optimization variables $\{\boldsymbol{t}_n\}_{n=1}^N \cup \{\boldsymbol{r}_m\}_{m=1}^M  \cup \boldsymbol{Q}$. Then, we solve three subproblems of (P1), which respectively optimize the transmit covariance matrix $\boldsymbol{Q}$, one transmit MA position $\boldsymbol{t}_n$, and one receive MA position $\boldsymbol{r}_m$, with all the other variables being fixed. The developed alternating optimization algorithm can obtain a (at least) locally optimal solution for (P1) by iteratively solving the above three subproblems in an alternate manner. After that, we derive more tractable expressions of the MIMO channel capacity for the asymptotically low-SNR regime, based on which we propose an alternative low-complexity solution for (P1). Finally, we consider the special cases of (P1) with a single MA at the transmitter/receiver, i.e., the SIMO/MISO setup, and propose simplified algorithms for them.

\subsection{Alternating Optimization for (P1)}
In this subsection, we introduce the framework of our proposed alternating optimization for solving (P1). The main idea is to iteratively solve a series of subproblems of (P1), each
aiming to optimize one single variable in $\{\boldsymbol{t}_n\}_{n=1}^N \cup \{\boldsymbol{r}_m\}_{m=1}^M  \cup \boldsymbol{Q}$ with the other $M+N$ variables being fixed, which tackles the main difficulty of the complicated coupling among the optimization variables.

\subsubsection{Optimization of $\boldsymbol{Q}$ with given $\{\boldsymbol{t}_n\}_{n=1}^N$ and $\{\boldsymbol{r}_m\}_{m=1}^M$}
In this subproblem, we aim to optimize the transmit covariance matrix $\boldsymbol{Q}$ with given transmit MA positions $\{\boldsymbol{t}_n\}_{n=1}^N$ and receive MA positions $\{\boldsymbol{r}_m\}_{m=1}^M$. Note that with given $\boldsymbol{H}(\tilde{\boldsymbol{t}}, \tilde{\boldsymbol{r}})$, (P1) is a convex optimization problem over $\boldsymbol{Q}$, and the optimal solution is given by the eigenmode transmission \cite{goldsmith2005wireless}. Specifically, denote  $\boldsymbol{H}(\tilde{\boldsymbol{t}}, \tilde{\boldsymbol{r}})=\tilde{\boldsymbol{U}} \tilde{\boldsymbol{\Lambda}} \tilde{\boldsymbol{V}}^H$  as the truncated singular value decomposition (SVD) of $\boldsymbol{H}(\tilde{\boldsymbol{t}}, \tilde{\boldsymbol{r}})$, with $\tilde{\boldsymbol{U}} \in \mathbb{C}^{M \times S}$, $\tilde{\boldsymbol{V}} \in \mathbb{C}^{N \times S}$, $\tilde{\boldsymbol{\Lambda}} \in \mathbb{C}^{S \times S}$, and  $S=\textrm{rank}(\boldsymbol{H}(\tilde{\boldsymbol{t}}, \tilde{\boldsymbol{r}})) \leq   \min \left(N, M\right)$  denoting the maximum number of data streams that can be transmitted over $\boldsymbol{H}(\tilde{\boldsymbol{t}}, \tilde{\boldsymbol{r}})$. The optimal $\boldsymbol{Q}$ is thus given by
\begin{equation}\label{Qstar}
  \boldsymbol{Q}^{\star}=\tilde{\boldsymbol{V}} \textrm{diag}([p_1^{\star}, p_2^{\star}, \ldots, p_S^{\star}]) \tilde{\boldsymbol{V}}^H,
\end{equation}
where $p_s^{\star}$ denotes the optimal power allocated to the  $s$th ($s=1,2,\ldots,S$) data stream following the water-filling strategy: $p_s^{\star}=\max \left(1/p_0-\sigma^2/\tilde{\boldsymbol{\Lambda}}[s,s]^2, 0\right)$, with $p_0$ satisfying $\sum_{s=1}^S p_s^{\star}=P$. Hence, the channel capacity with given  $\{\boldsymbol{t}_n\}_{n=1}^N$ and $\{\boldsymbol{r}_m\}_{m=1}^M$ can be expressed as
\begin{equation}
  C(\tilde{\boldsymbol{t}}, \tilde{\boldsymbol{r}})=\sum_{s=1}^S \log_2\left(1+\frac{\tilde{\boldsymbol{\Lambda}}[s,s]^2 p_s^{\star}}{\sigma^2}\right).
\end{equation}

\subsubsection{Optimization of $\boldsymbol{r}_m$ with given $\boldsymbol{Q}$, $\{\boldsymbol{r}_k, k\neq m\}_{k=1}^M$ and $\{\boldsymbol{t}_n\}_{n=1}^N$}
In this subproblem, we aim to optimize $\boldsymbol{r}_m$ in (P1) with given $\boldsymbol{Q}$, $\{\boldsymbol{t}_n\}_{n=1}^N$, and $\{\boldsymbol{r}_k, k\neq m\}_{k=1}^M$, $\forall m\in \mathcal{M}=\{1,2,\ldots,M\}$. Denote $\boldsymbol{Q}=\boldsymbol{U}_{Q} \boldsymbol{V}_{Q} \boldsymbol{U}_{Q}^H$  as the eigenvalue decomposition (EVD) of $\boldsymbol{Q}$, with $\boldsymbol{U}_{Q} \in \mathbb{C}^{N \times N}$ and $\boldsymbol{V}_{Q} \in \mathbb{C}^{N \times N}$. Note that since $\boldsymbol{Q}$ is a positive semi-definite matrix, all the diagonal elements in $\boldsymbol{V}_{Q}$ are non-negative real numbers. Given $\{\boldsymbol{t}_n\}_{n=1}^N$ and $\boldsymbol{Q}$, we define $\boldsymbol{W}(\tilde{\boldsymbol{r}})=\boldsymbol{H}(\tilde{\boldsymbol{t}}, \tilde{\boldsymbol{r}}) \boldsymbol{U}_{Q} \boldsymbol{V}_{Q}^{\frac{1}{2}} \in   \mathbb{C}^{M \times N}$ and denote the $m$th column vector of $\boldsymbol{W}(\tilde{\boldsymbol{r}})^H$ by $\boldsymbol{w}(\boldsymbol{r}_m) \in\mathbb{C}^{N}$, which is only determined by the position of the $m$th MA and can be written as
\begin{equation}\label{for3}
	\boldsymbol{w}(\boldsymbol{r}_m) = \boldsymbol{V}_{Q}^{\frac{1}{2}}  \boldsymbol{U}_{Q}^H   \boldsymbol{G}(\tilde{\boldsymbol{t}})^H \boldsymbol{\Sigma}^H    \boldsymbol{f}(\boldsymbol{r}_m).
\end{equation}
Thus, the objective function of (P1) can be rewritten in the following form with respect to $\tilde{\boldsymbol{r}}$:
\begin{align}\label{fW}
	b\left(\tilde{\boldsymbol{r}}\right) & \triangleq \log_{2} \det\left(\boldsymbol{I}_M+\frac{1}{\sigma^2} \boldsymbol{H}(\tilde{\boldsymbol{t}}, \tilde{\boldsymbol{r}}) \boldsymbol{Q} \boldsymbol{H}(\tilde{\boldsymbol{t}}, \tilde{\boldsymbol{r}})^H\right) \\
	&=\log_{2} \det\left(\boldsymbol{I}_M+\frac{1}{\sigma^{2}} \boldsymbol{W}(\tilde{\boldsymbol{r}}) \boldsymbol{W}(\tilde{\boldsymbol{r}})^H\right) \notag\\
	&\overset{(a)}=\log_{2} \det\left(\boldsymbol{I}_N+\frac{1}{\sigma^{2}} \boldsymbol{W}(\tilde{\boldsymbol{r}})^H \boldsymbol{W}(\tilde{\boldsymbol{r}})\right) \notag\\
	&=\log_{2} \det\left(\boldsymbol{I}_N+\frac{1}{\sigma^{2}} \sum_{m=1}^{M} \boldsymbol{w}(\boldsymbol{r}_m)\boldsymbol{w}(\boldsymbol{r}_m)^H\right), \notag
\end{align}
where the equality marked by $(a)$ holds due to $\det(\boldsymbol{I}_p+\boldsymbol{A}\boldsymbol{B}) = \det(\boldsymbol{I}_q+\boldsymbol{B}\boldsymbol{A})$ for $\boldsymbol{A} \in{\mathbb{C}^{p \times q}}$ and $\boldsymbol{B} \in{\mathbb{C}^{q \times p}}$. Notice from \eqref{fW} that $\boldsymbol{H}(\tilde{\boldsymbol{t}}, \tilde{\boldsymbol{r}}) \boldsymbol{Q} \boldsymbol{H}(\tilde{\boldsymbol{t}}, \tilde{\boldsymbol{r}})^H$ is in fact the
summation of $M$ rank-one matrices, which is a unique structure of MA-enabled MIMO channel and implies that $\{\boldsymbol{r}_m\}_{m=1}^M$ should be designed to achieve an optimal balance between the $M$ matrices for maximizing the channel capacity. Note that the equivalent objective function of (P1) shown in \eqref{fW} is in an explicit form which decouples the position variables for all the $M$ MAs, i.e., $\{\boldsymbol{r}_m\}_{m=1}^M$, which facilitates the following optimization of $\boldsymbol{r}_m$.

Remove $\boldsymbol{w}(\boldsymbol{r}_m)$ from $\boldsymbol{W}(\tilde{\boldsymbol{r}})^H$ and denote the remaining $N\times (M-1)$ sub-matrix by $\boldsymbol{W}_m^H = [\boldsymbol{w}(\boldsymbol{r}_1), \boldsymbol{w}(\boldsymbol{r}_2), \ldots, \boldsymbol{w}(\boldsymbol{r}_{m-1}), \boldsymbol{w}(\boldsymbol{r}_{m+1}), \ldots, \boldsymbol{w}(\boldsymbol{r}_M)]$. Thus, the objective function of (P1) in \eqref{fW} can be rewritten as \cite{gharavi2004fast}
\begin{align}\label{fr}
	&\bar{b}(\boldsymbol{r}_m) = \log_{2} \det\Big(\boldsymbol{I}_N+\frac{1}{\sigma^{2}} \Big(\boldsymbol{W}_m^H\boldsymbol{W}_m + \boldsymbol{w}(\boldsymbol{r}_m)\boldsymbol{w}(\boldsymbol{r}_m)^H\Big)\Big) \notag\\
	&\overset{(b_1)}= \log_{2} \det\Big(\boldsymbol{I}_N  \!  + \!  \frac{1}{\sigma^{2}} \Big( \boldsymbol{I}_N  \! + \!  \frac{1}{\sigma^{2}} \boldsymbol{W}_m^H\boldsymbol{W}_m \Big)^{\!-1} \! \! \!  \boldsymbol{w}(\boldsymbol{r}_m) \boldsymbol{w}(\boldsymbol{r}_m)^H  \Big)  \notag \\
	& ~~~~ +\log_{2} \det\Big(\boldsymbol{I}_N+\frac{1}{\sigma^{2}} \boldsymbol{W}_m^H\boldsymbol{W}_m\Big) \notag \\
	&\overset{(b_2)}= \log_{2} \Big(1+\frac{1}{\sigma^{2}} \boldsymbol{w}(\boldsymbol{r}_m)^H \Big( \boldsymbol{I}_N+\frac{1}{\sigma^{2}} \boldsymbol{W}_m^H\boldsymbol{W}_m \Big)^{-1} \boldsymbol{w}(\boldsymbol{r}_m) \Big)  \notag \\
	& ~~~~ +\log_{2} \det\Big(\boldsymbol{I}_N+\frac{1}{\sigma^{2}} \boldsymbol{W}_m^H\boldsymbol{W}_m\Big),
\end{align}
where the equality marked by $(b_1)$ holds due to the fact that $\det(\boldsymbol{A}\boldsymbol{B})=\det(\boldsymbol{A})\det(\boldsymbol{B})$ holds for two equal-sized square matrices $\boldsymbol{A}$ and $\boldsymbol{B}$. The equality marked by $(b_2)$ holds because of $\det(\boldsymbol{I}_p+\boldsymbol{A}\boldsymbol{B}) = \det(\boldsymbol{I}_q+\boldsymbol{B}\boldsymbol{A})$ for $\boldsymbol{A} \in{\mathbb{C}^{p \times q}}$ and $\boldsymbol{B} \in{\mathbb{C}^{q \times p}}$. Notice from \eqref{fr} that the maximization of $\bar{b}(\boldsymbol{r}_m)$ for given $\boldsymbol{Q}$ and $\{\boldsymbol{r}_k, k\neq m\}_{k=1}^M$ is equivalent to maximizing
\begin{equation}\label{gr}
  g(\boldsymbol{r}_m) \triangleq \boldsymbol{w}(\boldsymbol{r}_m)^H \boldsymbol{A}_m \boldsymbol{w}(\boldsymbol{r}_m),
\end{equation}
where $\boldsymbol{A}_m \triangleq \left( \boldsymbol{I}_N+\frac{1}{\sigma^{2}} \boldsymbol{W}_m^H\boldsymbol{W}_m \right)^{-1} \in{\mathbb{C}^{N\times{N}}}$ is a positive definite matrix independent to $\boldsymbol{r}_m$. During the iterations, to reduce the computational complexity of matrix inverse operation for $2 \leq m \leq M$, matrix $\boldsymbol{A}_m$ can be updated based on $\boldsymbol{A}_{m-1}$. Specifically, we define $\boldsymbol{Z}_1=[\boldsymbol{w}(\boldsymbol{r}_{m-1}),\boldsymbol{w}(\boldsymbol{r}_m)]\in{\mathbb{C}^{N\times{2}}}$ and $\boldsymbol{Z}_2=[\boldsymbol{w}(\boldsymbol{r}_{m-1}),-\boldsymbol{w}(\boldsymbol{r}_m)]\in{\mathbb{C}^{N\times{2}}}$ such that
\begin{equation}
  \boldsymbol{W}_m^H\boldsymbol{W}_m = \boldsymbol{W}_{m-1}^H\boldsymbol{W}_{m-1} + \boldsymbol{Z}_1 \boldsymbol{Z}_2^H.
\end{equation}
According to the matrix inversion lemma, we have \cite{petersen2006matrix}
\begin{align}\label{Am}
  \boldsymbol{A}_m = \boldsymbol{A}_{m-1}-\frac{1}{\sigma^{2}}\boldsymbol{A}_{m-1} \boldsymbol{Z}_1\left(\boldsymbol{I}_2+\frac{1}{\sigma^{2}}\boldsymbol{Z}_2^H \boldsymbol{A}_{m-1} \boldsymbol{Z}_1\right)^{-1} \boldsymbol{Z}_2^H \boldsymbol{A}_{m-1}.
\end{align}
Furthermore, we define a positive definite matrix
\begin{equation}\label{Bm}
	\boldsymbol{B}_m \triangleq \boldsymbol{\Sigma} \boldsymbol{G}(\tilde{\boldsymbol{t}}) \boldsymbol{U}_{Q} \boldsymbol{V}_{Q}^{\frac{1}{2}}         \boldsymbol{A}_m \boldsymbol{V}_{Q}^{\frac{1}{2}}  \boldsymbol{U}_{Q}^H   \boldsymbol{G}(\tilde{\boldsymbol{t}})^H \boldsymbol{\Sigma}^H \in{\mathbb{C}^{L_r\times{L_r}}}.
\end{equation}
According to \eqref{for3}, $g(\boldsymbol{r}_m)$ in \eqref{gr} can be rewritten in the following form with respect to $\boldsymbol{f}(\boldsymbol{r}_m)$:
\begin{align}\label{gr2}
  g(\boldsymbol{r}_m) &= \boldsymbol{f}(\boldsymbol{r}_m)^H \boldsymbol{\Sigma} \boldsymbol{G}(\tilde{\boldsymbol{t}}) \boldsymbol{U}_{Q} \boldsymbol{V}_{Q}^{\frac{1}{2}}      \boldsymbol{A}_m \boldsymbol{V}_{Q}^{\frac{1}{2}}  \boldsymbol{U}_{Q}^H   \boldsymbol{G}(\tilde{\boldsymbol{t}})^H \boldsymbol{\Sigma}^H    \boldsymbol{f}(\boldsymbol{r}_m) \\
  &= \boldsymbol{f}(\boldsymbol{r}_m)^H \boldsymbol{B}_m \boldsymbol{f}(\boldsymbol{r}_m). \notag
\end{align}
Notice that $\boldsymbol{B}_m$ is a constant matrix independent to $\boldsymbol{r}_m$. As a result, the subproblem for optimizing $\boldsymbol{r}_m$ can be expressed as
\begin{subequations}
\begin{align}
\textrm {(P2-m)}~~\max_{\boldsymbol{r}_m} \quad & \boldsymbol{f}(\boldsymbol{r}_m)^H \boldsymbol{B}_m \boldsymbol{f}(\boldsymbol{r}_m) \label{P2a}\\
\text{s.t.} \quad & \boldsymbol{r}_m \in \mathcal{C}_r, \label{P2b}\\
                  & \|\boldsymbol{r}_m-\boldsymbol{r}_k\|_2 \geq D,~~ k = 1,2,\ldots,M,~~ k\neq m.\label{P2c}
\end{align}
\end{subequations}
As can be observed, the objective function of (P2-m) is a non-concave function over $\boldsymbol{r}_m$, and the minimum distance constraints \eqref{P2c} are also non-convex. Thus, problem (P2-m) is still a non-convex optimization problem which is difficult to solve. We will present the solution for (P2-m) in Section III-B, where the successive convex approximation (SCA) is used for relaxing this problem.

\subsubsection{Optimization of $\boldsymbol{t}_n$ with given $\boldsymbol{Q}$, $\{\boldsymbol{t}_l, l\neq n\}_{l=1}^N$ and $\{\boldsymbol{r}_m\}_{m=1}^M$}
In this subproblem, we aim to optimize $\boldsymbol{t}_n$ in (P1) with given $\boldsymbol{Q}$, $\{\boldsymbol{r}_m\}_{m=1}^M$, and $\{\boldsymbol{t}_l, l\neq n\}_{l=1}^N$, $\forall n\in \mathcal{N}=\{1,2,\ldots,N\}$. Since the singular values of channel matrices $\boldsymbol{H}(\tilde{\boldsymbol{t}}, \tilde{\boldsymbol{r}})$ and $\boldsymbol{H}(\tilde{\boldsymbol{t}}, \tilde{\boldsymbol{r}})^H$ are the same, the channel capacity of $\boldsymbol{H}(\tilde{\boldsymbol{t}}, \tilde{\boldsymbol{r}})$ is always equal to that of $\boldsymbol{H}(\tilde{\boldsymbol{t}}, \tilde{\boldsymbol{r}})^H$ under the same transmit power constraint, which is also known as the channel reciprocity for MIMO systems, i.e.,
\begin{align}
  &\max_{\boldsymbol{Q}: {\rm {Tr}}(\boldsymbol{Q}) \leq P, \boldsymbol{Q} \succeq \mathbf{0}}\log_{2} \det\left(\boldsymbol{I}_M+\frac{1}{\sigma^2} \boldsymbol{H}(\tilde{\boldsymbol{t}}, \tilde{\boldsymbol{r}}) \boldsymbol{Q} \boldsymbol{H}(\tilde{\boldsymbol{t}}, \tilde{\boldsymbol{r}})^H\right) \\
  =& \max_{\boldsymbol{S}: {\rm {Tr}}(\boldsymbol{S}) \leq P, \boldsymbol{S} \succeq \mathbf{0}}\log_{2} \det\left(\boldsymbol{I}_N+\frac{1}{\sigma^2} \boldsymbol{H}(\tilde{\boldsymbol{t}}, \tilde{\boldsymbol{r}})^H \boldsymbol{S} \boldsymbol{H}(\tilde{\boldsymbol{t}}, \tilde{\boldsymbol{r}})\right), \notag
\end{align}
where $\boldsymbol{S}\in \mathbb{C}^{M \times M}$ is the equivalent covariance matrix of transmit signals for $\boldsymbol{H}(\tilde{\boldsymbol{t}}, \tilde{\boldsymbol{r}})^H$, with ${\rm {Tr}}(\boldsymbol{S}) \leq P$ and $\boldsymbol{S} \succeq \mathbf{0}$. Similar to the optimization of $\boldsymbol{Q}$ with given $\{\boldsymbol{t}_n\}_{n=1}^N$ and $\{\boldsymbol{r}_m\}_{m=1}^M$, the optimal $\boldsymbol{S}$ is given by the eigenmode transmission. Specifically, the truncated SVD of $\boldsymbol{H}(\tilde{\boldsymbol{t}}, \tilde{\boldsymbol{r}})^H$ is written as $\boldsymbol{H}(\tilde{\boldsymbol{t}}, \tilde{\boldsymbol{r}})^H= \tilde{\boldsymbol{V}} \tilde{\boldsymbol{\Lambda}} \tilde{\boldsymbol{U}}^H$. Then, the optimal $\boldsymbol{S}$ is given by
\begin{equation}\label{Sstar}
  \boldsymbol{S}^{\star}=\tilde{\boldsymbol{U}} \textrm{diag}([p_1^{\star}, p_2^{\star}, \ldots, p_S^{\star}]) \tilde{\boldsymbol{U}}^H,
\end{equation}
where the optimal power allocated to $S$ date streams, $\{p_s^{\star}\}_{s=1}^S$, is given by \eqref{Qstar}.

Similar to the procedure for optimizing $\boldsymbol{r}_m$, we denote $\boldsymbol{S}=\boldsymbol{U}_{S} \boldsymbol{V}_{S} \boldsymbol{U}_{S}^H$  as the EVD of $\boldsymbol{S}$, with $\boldsymbol{U}_{S} \in \mathbb{C}^{M \times M}$ and $\boldsymbol{V}_{S} \in \mathbb{C}^{M \times M}$. Since $\boldsymbol{S}$ is a positive semi-definite matrix, all the diagonal elements in $\boldsymbol{V}_{S}$ are non-negative real numbers. Based on this, we define $\boldsymbol{P}(\tilde{\boldsymbol{t}})=\boldsymbol{H}(\tilde{\boldsymbol{t}}, \tilde{\boldsymbol{r}})^H \boldsymbol{U}_{S} \boldsymbol{V}_{S}^{\frac{1}{2}} \in   \mathbb{C}^{N \times M}$ and denote the $n$th column vector of $\boldsymbol{P}(\tilde{\boldsymbol{t}})^H$ by $\boldsymbol{p}(\boldsymbol{t}_n) = \boldsymbol{V}_{S}^{\frac{1}{2}}  \boldsymbol{U}_{S}^H   \boldsymbol{F}(\tilde{\boldsymbol{r}})^H \boldsymbol{\Sigma}    \boldsymbol{g}\left(\boldsymbol{t}_n\right) \in\mathbb{C}^{M}$. Then, remove $\boldsymbol{p}(\boldsymbol{t}_n)$ from $\boldsymbol{P}(\tilde{\boldsymbol{t}})^H$ and denote the remaining $M\times (N-1)$ sub-matrix by $\boldsymbol{P}_n^H = [\boldsymbol{p}(\boldsymbol{t}_1), \boldsymbol{p}(\boldsymbol{t}_2), \ldots, \boldsymbol{p}(\boldsymbol{t}_{n-1}), \boldsymbol{p}(\boldsymbol{t}_{n+1}), \ldots, \boldsymbol{p}(\boldsymbol{t}_N)]$. Let $\boldsymbol{C}_n \triangleq\left( \boldsymbol{I}_M+\frac{1}{\sigma^{2}} \boldsymbol{P}_n^H\boldsymbol{P}_n \right)^{-1} \in{\mathbb{C}^{M\times{M}}}$, which can be updated similarly to \eqref{Am}. Next, we define
\begin{equation}\label{Dn}
	\boldsymbol{D}_n \triangleq \boldsymbol{\Sigma}^H \boldsymbol{F}(\tilde{\boldsymbol{r}}) \boldsymbol{U}_{S} \boldsymbol{V}_{S}^{\frac{1}{2}} \boldsymbol{C}_n \boldsymbol{V}_{S}^{\frac{1}{2}} \boldsymbol{U}_{S}^H \boldsymbol{F}(\tilde{\boldsymbol{r}})^H \boldsymbol{\Sigma} \in{\mathbb{C}^{L_t\times{L_t}}}.
\end{equation}
As a result, the optimization of $\boldsymbol{t}_n$ can be expressed as
\begin{subequations}
\begin{align}
\textrm {(P3-n)}~~\max_{\boldsymbol{t}_n} \quad & \boldsymbol{g}\left(\boldsymbol{t}_n\right)^H \boldsymbol{D}_n \boldsymbol{g}\left(\boldsymbol{t}_n\right) \label{P3a}\\
\text{s.t.} \quad & \boldsymbol{t}_n \in \mathcal{C}_t, \label{P3b}\\
                  & \|\boldsymbol{t}_n-\boldsymbol{t}_l\|_2 \geq D,~~ l = 1,2,\ldots,N,~~ l\neq n.\label{P3c}
\end{align}
\end{subequations}
%\begin{align}\label{ft}
%\tilde{f}\left(\boldsymbol{t}_n\right) &=\log_{2} \det\left(\boldsymbol{I}_N+\frac{1}{\sigma^2} \boldsymbol{H}(\tilde{\boldsymbol{t}}, \tilde{\boldsymbol{r}})^H \boldsymbol{S} \boldsymbol{H}(\tilde{\boldsymbol{t}}, \tilde{\boldsymbol{r}})\right) \\
%&=\log_{2} \det\left(\boldsymbol{I}_N+\frac{1}{\sigma^2} \boldsymbol{P}(\tilde{\boldsymbol{t}}) \boldsymbol{P}(\tilde{\boldsymbol{t}})^H\right) \notag\\
%&=\log_{2} \det\left(\boldsymbol{I}_M+\frac{1}{\sigma^2} \boldsymbol{P}(\tilde{\boldsymbol{t}})^H \boldsymbol{P}(\tilde{\boldsymbol{t}})\right) \notag\\
%& = \log_{2} \det\left(\boldsymbol{I}_M+\frac{1}{\sigma^{2}} \left(\boldsymbol{P}_n^H\boldsymbol{P}_n + \boldsymbol{p}(\boldsymbol{t}_n)\boldsymbol{p}(\boldsymbol{t}_n)^H\right)\right) \notag \\
%&= \log_{2} \left(1+\frac{1}{\sigma^{2}} \boldsymbol{p}(\boldsymbol{t}_n)^H \left( \boldsymbol{I}_M+\frac{1}{\sigma^{2}} \boldsymbol{P}_n^H\boldsymbol{P}_n \right)^{-1} \boldsymbol{p}(\boldsymbol{t}_n) \right) + \notag \\
%& ~~~~ \log_{2} \det\left(\boldsymbol{I}_M+\frac{1}{\sigma^{2}} \boldsymbol{P}_n^H\boldsymbol{P}_n\right) \notag
%\end{align}
As can be observed, problem (P3-n) has the same structure as (P2-m). Thus, in the following, we only provide the details of the solution for problem (P2-m), and problem (P3-n) can be solved in a similar way.

\subsection{Solution for Problem (P2-m)}
To solve problem (P2-m), we adopt the
SCA technique to optimize the position of the $m$th receive MA, $\boldsymbol{r}_m$. Although the objective function $\boldsymbol{f}(\boldsymbol{r}_m)^H \boldsymbol{B}_m \boldsymbol{f}(\boldsymbol{r}_m)$ is a non-concave function over $\boldsymbol{r}_m$, it is convex with respect to $\boldsymbol{f}(\boldsymbol{r}_m)$. Recall that any convex function is globally lower-bounded by its first-order Taylor expansion at any point. With given local point $\boldsymbol{r}_m^i$ in the $i$th iteration of SCA, we obtain the following lower
bound on $\boldsymbol{f}(\boldsymbol{r}_m)^H \boldsymbol{B}_m \boldsymbol{f}(\boldsymbol{r}_m)$ as \cite{wu2018joint}
\begin{align}\label{gr3}
  g(\boldsymbol{r}_m) &= \boldsymbol{f}(\boldsymbol{r}_m)^H \boldsymbol{B}_m \boldsymbol{f}(\boldsymbol{r}_m) \\
  &\geq \boldsymbol{f}(\boldsymbol{r}_m^i)^H \boldsymbol{B}_m \boldsymbol{f}(\boldsymbol{r}_m^i) + 2{\rm{Re}}\left\{\boldsymbol{f}(\boldsymbol{r}_m^i)^H \boldsymbol{B}_m \left(\boldsymbol{f}(\boldsymbol{r}_m) - \boldsymbol{f}(\boldsymbol{r}_m^i)\right)\right\} \notag\\
  &=2\underbrace{{\rm{Re}}\left\{\boldsymbol{f}(\boldsymbol{r}_m^i)^H \boldsymbol{B}_m \boldsymbol{f}(\boldsymbol{r}_m)\right\}}_{\bar{g}(\boldsymbol{r}_m)} - \underbrace{\boldsymbol{f}(\boldsymbol{r}_m^i)^H \boldsymbol{B}_m  \boldsymbol{f}(\boldsymbol{r}_m^i)}_{\textrm{constant}}, \notag
\end{align}
where $\boldsymbol{r}_m^i \in{\mathbb{R}^{2}}$ is a constant vector which represents the value of $\boldsymbol{r}_m$ in the $i$th iteration. Thus, maximizing $\boldsymbol{f}(\boldsymbol{r}_m)^H \boldsymbol{B}_m \boldsymbol{f}(\boldsymbol{r}_m)$ can be simplified to maximizing $\bar{g}(\boldsymbol{r}_m) \triangleq {\rm{Re}}\left\{\boldsymbol{f}(\boldsymbol{r}_m^i)^H \boldsymbol{B}_m \boldsymbol{f}(\boldsymbol{r}_m)\right\}$. Although $\bar{g}(\boldsymbol{r}_m)$ is a linear function over $\boldsymbol{f}(\boldsymbol{r}_m)$, it is still neither concave nor convex over $\boldsymbol{r}_m$. Thus, we cannot construct the surrogate function lower-bounding the objective function only by the first-order Taylor expansion of $\bar{g}(\boldsymbol{r}_m)$. Alternatively, we construct a surrogate function  that locally approximates the objective function by using the second-order Taylor expansion. Denote the gradient vector and Hessian matrix of $\bar{g}(\boldsymbol{r}_m)$ over $\boldsymbol{r}_m$ by $\nabla \bar{g}(\boldsymbol{r}_m) \in{\mathbb{R}^{2}}$ and $\nabla^2 \bar{g}(\boldsymbol{r}_m) \in{\mathbb{R}^{2\times{2}}}$, respectively, with their derivations provided in Appendix A. Then, we construct a positive real number $\delta_m$ making $\delta_m\boldsymbol{I}_2 \succeq \nabla^2 \bar{g}(\boldsymbol{r}_m)$, with the closed-form expression given in Appendix B. Thus, based on Taylor’s theorem, we can find a quadratic surrogate function to globally lower-bound the objective function $\bar{g}(\boldsymbol{r}_m)$ as \cite{magnus1995matrix}
\begin{align}\label{gr4}
  \bar{g}(\boldsymbol{r}_m) &\geq \bar{g}(\boldsymbol{r}_m^i) + \nabla \bar{g}(\boldsymbol{r}_m^i)^T \left(\boldsymbol{r}_m-\boldsymbol{r}_m^i\right) - \frac{\delta_m}{2}\left(\boldsymbol{r}_m-\boldsymbol{r}_m^i\right)^T \left(\boldsymbol{r}_m-\boldsymbol{r}_m^i\right) \\
  &= \underbrace{-\frac{\delta_m}{2}\boldsymbol{r}_m^T \boldsymbol{r}_m + \left( \nabla \bar{g}(\boldsymbol{r}_m^i) + \delta_m\boldsymbol{r}_m^i \right)^T \boldsymbol{r}_m}_{\tilde{g}(\boldsymbol{r}_m)} +   \underbrace{\bar{g}(\boldsymbol{r}_m^i) - \frac{\delta_m}{2}(\boldsymbol{r}_m^i)^T \boldsymbol{r}_m^i}_{\textrm{constant}}. \notag
\end{align}
Thus, maximizing $\bar{g}(\boldsymbol{r}_m)$ can be converted to maximizing $\tilde{g}(\boldsymbol{r}_m) \triangleq -\frac{\delta_m}{2}\boldsymbol{r}_m^T \boldsymbol{r}_m + ( \nabla \bar{g}(\boldsymbol{r}_m^i) + \delta_m\boldsymbol{r}_m^i )^T \boldsymbol{r}_m$. To this end, in the $i$th iteration of SCA, the optimization problem of the $m$th receive MA position $\boldsymbol{r}_m$ can be relaxed as
\begin{align}\label{P4}
\textrm {(P4-m)}~~\max_{\boldsymbol{r}_m} \quad & -\frac{\delta_m}{2}\boldsymbol{r}_m^T \boldsymbol{r}_m + \left( \nabla \bar{g}(\boldsymbol{r}_m^i) + \delta_m\boldsymbol{r}_m^i \right)^T \boldsymbol{r}_m \\
\text{s.t.} \quad & \eqref{P2b}, \eqref{P2c}. \notag
\end{align}
Since the objective function of (P4-m) is a concave quadratic function over $\boldsymbol{r}_m$, the global optimum for maximizing $\tilde{g}(\boldsymbol{r}_m)$ by neglecting constraints $\eqref{P2b}$ and $\eqref{P2c}$ can be obtained in closed form as follows
\begin{equation}\label{rstar}
  \boldsymbol{r}_{m,i+1}^{\star} =  \frac{1}{\delta_m} \nabla \bar{g}(\boldsymbol{r}_m^i) + \boldsymbol{r}_m^i.
\end{equation}
If $\boldsymbol{r}_{m,i+1}^{\star}$ satisfies \eqref{P2b} and \eqref{P2c}, it is the global optimum for problem (P4-m). However, if $\boldsymbol{r}_{m,i+1}^{\star}$ does not satisfy \eqref{P2b} or \eqref{P2c}, we cannot obtain the optimal solution due to the non-convex constraints \eqref{P2c}. In this case, we propose the following solution. Denote the gradient vector of $\|\boldsymbol{r}_m-\boldsymbol{r}_k\|_2$ over $\boldsymbol{r}_m$ as $\nabla \|\boldsymbol{r}_m-\boldsymbol{r}_k\|_2 = (\boldsymbol{r}_m-\boldsymbol{r}_k)/\|\boldsymbol{r}_m-\boldsymbol{r}_k\|_2$. Since the denominator term $\|\boldsymbol{r}_m-\boldsymbol{r}_k\|_2 \geq D >0$, the gradient vector always exists. As $\|\boldsymbol{r}_m-\boldsymbol{r}_k\|_2$ is a convex function with respect to $\boldsymbol{r}_m$, we have the following inequality by applying the first-order Taylor expansion at the given point $\boldsymbol{r}_m^i$:
\begin{align}
  \|\boldsymbol{r}_m-\boldsymbol{r}_k\|_2 &\geq \|\boldsymbol{r}_m^i-\boldsymbol{r}_k\|_2 + \left(\nabla \|\boldsymbol{r}_m^i-\boldsymbol{r}_k\|_2\right)^T (\boldsymbol{r}_m-\boldsymbol{r}_m^i) \\
  &=\|\boldsymbol{r}_m^i-\boldsymbol{r}_k\|_2 + \frac{1}{\|\boldsymbol{r}_m^i-\boldsymbol{r}_k\|_2}(\boldsymbol{r}_m^i-\boldsymbol{r}_k)^T (\boldsymbol{r}_m-\boldsymbol{r}_m^i) \notag\\
  &=\frac{1}{\|\boldsymbol{r}_m^i-\boldsymbol{r}_k\|_2}(\boldsymbol{r}_m^i-\boldsymbol{r}_k)^T (\boldsymbol{r}_m-\boldsymbol{r}_k). \notag
\end{align}
Hereto, for a given $\boldsymbol{r}_m^i$ obtained in the $i$th iteration, if $\boldsymbol{r}_{m,i+1}^{\star}$ does not satisfy \eqref{P2b} or \eqref{P2c}, the convex position optimization problem for the $m$th receive MA is transformed into
\begin{subequations}\label{P5}
\begin{align}
\textrm {(P5-m)}~~\max_{\boldsymbol{r}_m} \quad & -\frac{\delta_m}{2}\boldsymbol{r}_m^T \boldsymbol{r}_m + \left( \nabla \bar{g}(\boldsymbol{r}_m^i) + \delta_m\boldsymbol{r}_m^i \right)^T \boldsymbol{r}_m \label{P5a}\\
\text{s.t.} \quad & \frac{1}{\|\boldsymbol{r}_m^i-\boldsymbol{r}_k\|_2}(\boldsymbol{r}_m^i-\boldsymbol{r}_k)^T (\boldsymbol{r}_m-\boldsymbol{r}_k) \geq D,~~ k = 1,2,\ldots,M,~~ k\neq m, \label{P5b}\\
                  & \eqref{P2b}. \notag
\end{align}
\end{subequations}
Since the objective function \eqref{P5a} is quadratic and constraints \eqref{P5b} are linear with respect to $\boldsymbol{r}_m$, (P5-m) is a quadratic programming (QP) problem and can be efficiently solved by using quadprog \cite{turlach2007quad}. The details of the proposed algorithm to solve problem (P2-m) are summarized in Algorithm~\ref{alg1}. Specifically, from step 4 to step 6, we compute the gradient vector and Hessian matrix of $\bar{g}(\boldsymbol{r}_m)$ at $\boldsymbol{r}_m^i$ via \eqref{gradg} and \eqref{grad2g}, respectively, and then obtain $\delta_m$ via \eqref{delta}. Then, we obtain $\boldsymbol{r}_{m,i+1}^{\star}$ as the global optimum to maximize $\tilde{g}(\boldsymbol{r}_m)$ in step 7. If $\boldsymbol{r}_{m,i+1}^{\star}$ satisfies \eqref{P2b} and \eqref{P2c}, we set it as $\boldsymbol{r}_m^{i+1}$ for the next iteration. Otherwise we solve the QP problem \eqref{P5} to obtain $\boldsymbol{r}_m^{i+1}$ in step 11. The proposed algorithm terminates when the increase of the objective value in \eqref{P2a} is below $\epsilon_1$, which is a predefined convergence threshold. Finally, we output the position of the $m$th receive MA in step 15.

\begin{algorithm}[!t]
	\caption{ SCA for Solving Problem (P2-m)}
	\label{alg1}
	\begin{algorithmic}[1]
		\STATE \emph{Input:} $M$, $L_r$, $\{\boldsymbol{r}_k, k\neq m\}_{k=1}^M$, $\boldsymbol{B}_m$, $\{\theta_r^q\}_{q=1}^{L_r}$, $\{\phi_r^q\}_{q=1}^{L_r}$, $\mathcal{C}_r$, $D$, $\epsilon_1$, $\boldsymbol{r}_m^0$.
        \STATE Initialization:  $i \leftarrow 0$.
        \WHILE{Increase of the objective value in \eqref{P2a} is above $\epsilon_1$}
        \STATE Obtain $\boldsymbol{b}$ via \eqref{b}.
        \STATE Compute $\nabla \bar{g}(\boldsymbol{r}_m^i)$ and $\nabla^2 \bar{g}(\boldsymbol{r}_m^i)$ via \eqref{gradg} and \eqref{grad2g}, respectively.
        \STATE Update $\delta_m$ via \eqref{delta}.
        \STATE Obtain $\boldsymbol{r}_{m,i+1}^{\star}$ via \eqref{rstar}.
        \IF{$\boldsymbol{r}_{m,i+1}^{\star}$ satisfies \eqref{P2b} and \eqref{P2c}}
            \STATE $\boldsymbol{r}_m^{i+1} \leftarrow \boldsymbol{r}_{m,i+1}^{\star}$.
        \ELSE
            \STATE Obtain $\boldsymbol{r}_m^{i+1}$ by solving \eqref{P5}.
        \ENDIF
        \STATE $i \leftarrow i+1$.
        \ENDWHILE
        \STATE \emph{Output:} $\boldsymbol{r}_m$.
	\end{algorithmic}
\end{algorithm}

Next we analyze the convergence of the proposed Algorithm~\ref{alg1}. Denote the two constant terms in \eqref{gr3} and \eqref{gr4} by $\Gamma_1(\boldsymbol{r}_m^i) = -\boldsymbol{f}(\boldsymbol{r}_m^i)^H \boldsymbol{B}_m  \boldsymbol{f}(\boldsymbol{r}_m^i)$ and $\Gamma_2(\boldsymbol{r}_m^i) = \bar{g}(\boldsymbol{r}_m^i) - \frac{\delta_m}{2}(\boldsymbol{r}_m^i)^T \boldsymbol{r}_m^i$, respectively. Then, for the $i$th iteration, the objective function in \eqref{P2a} can be written as
\begin{align}\label{convergence}
  g(\boldsymbol{r}_m^i) &\overset{(c_1)}= 2\left(\tilde{g}(\boldsymbol{r}_m^i) +  \Gamma_2(\boldsymbol{r}_m^i)\right) + \Gamma_1(\boldsymbol{r}_m^i) \\
  &\overset{(c_2)}\leq 2\left(\tilde{g}(\boldsymbol{r}_m^{i+1}) +  \Gamma_2(\boldsymbol{r}_m^i)\right) + \Gamma_1(\boldsymbol{r}_m^i) \overset{(c_3)}\leq 2\bar{g}(\boldsymbol{r}_m^{i+1}) + \Gamma_1(\boldsymbol{r}_m^i) 
  \overset{(c_4)}\leq g(\boldsymbol{r}_m^{i+1}),\notag
\end{align}
where the equality marked by $(c_1)$ holds because the first-order Taylor expansion in \eqref{gr3} and the second-order Taylor expansion in \eqref{gr4} are tight at $\boldsymbol{r}_m^i$. The inequality marked by $(c_2)$ holds because we maximize the value of $\tilde{g}(\boldsymbol{r}_m)$ in the $i$th iteration, and the equality can be achieved by choosing $\boldsymbol{r}_m^{i+1}=\boldsymbol{r}_m^i$. The inequalities marked by $(c_3)$ and $(c_4)$ hold because $\tilde{g}(\boldsymbol{r}_m) + \Gamma_2(\boldsymbol{r}_m^i)$ and $2\bar{g}(\boldsymbol{r}_m) + \Gamma_1(\boldsymbol{r}_m^i)$ are lower bounds on $\bar{g}(\boldsymbol{r}_m)$ and $g(\boldsymbol{r}_m)$ at $\boldsymbol{r}_m^{i+1}$, respectively. Thus, sequence $\{g(\boldsymbol{r}_m^i)\}_{i=0}^{\infty}$ is non-decreasing and will converge to a maximum value.

In the following, we analyze the computational complexity of Algorithm~\ref{alg1}. Specifically, from step 4 to step 7, the complexities of calculating $\boldsymbol{b}$, $\nabla \bar{g}(\boldsymbol{r}_m^i)$, $\nabla^2 \bar{g}(\boldsymbol{r}_m^i)$, $\delta_m$, and $\boldsymbol{r}_{m,i+1}^{\star}$ are $\mathcal{O}(N L_r)$, $\mathcal{O}(L_r)$, $\mathcal{O}(L_r)$, $\mathcal{O}(1)$, and $\mathcal{O}(1)$, respectively. In step 11, the complexity to solve the QP problem (P5-m) is $\mathcal{O}(M^{1.5}\ln(1/\beta))$ with accuracy $\beta$ for the interior-point method \cite{ben2001lecture}. Let $\gamma_r^1$ and $\gamma_r^2$ denote the maximum number of inner iterations (i.e., the number of times that steps 4-13 are repeated) and the maximum number of iterations required to perform step 11, respectively. Thus, the total complexity of Algorithm~\ref{alg1} is $\mathcal{O}(N L_r \gamma_r^1 + M^{1.5}\ln(1/\beta)\gamma_r^2)$, which is polynomial over $N$, $L_r$, and $M$.

\subsection{Solution for Problem (P3-n)}
Since problem (P3-n) has a similar structure as (P2-m), we can modify Algorithm~\ref{alg1} to obtain the position of the $n$th transmit MA $\boldsymbol{t}_n$ by replacing
\begin{align}
  \left\{ M, L_r, \boldsymbol{r}_m, \boldsymbol{B}_m, \mathcal{C}_r, \{\boldsymbol{r}_k, k\neq m\}_{k=1}^M, \{\theta_r^q\}_{q=1}^{L_r}, \{\phi_r^q\}_{q=1}^{L_r}  \right\} \notag
\end{align}
with
\begin{align}
  \left\{ N, L_t, \boldsymbol{t}_n, \boldsymbol{D}_n, \mathcal{C}_t, \{\boldsymbol{t}_l, l\neq n\}_{l=1}^N, \{\theta_t^p\}_{p=1}^{L_t}, \{\phi_t^p\}_{p=1}^{L_t}  \right\}. \notag
\end{align}

Similarly, the monotonic convergence is guaranteed for solving problem (P3-n) with Algorithm~\ref{alg1}. The corresponding computational complexity is $\mathcal{O}(M L_t \gamma_t^1 + N^{1.5}\ln(1/\beta)\gamma_t^2)$, with $\gamma_t^1$ and $\gamma_t^2$ denoting the maximum number of iterations to perform steps 4-13 and step 11, respectively.

\subsection{Overall Algorithm}
With the solutions for problem (P2-m) and (P3-n) derived above, we are ready to complete our proposed alternating optimization algorithm for solving (P1). The overall algorithm is summarized in Algorithm~\ref{alg2}. Specifically, in step 4, we first obtain the optimal solution of $\boldsymbol{Q}$ for (P1) with given $\{\boldsymbol{r}_m\}_{m=1}^M$ and $\{\boldsymbol{t}_n\}_{n=1}^N$ shown in \eqref{Qstar}. Then, from step 5 to step 8, we optimize the positions of $M$ receive MAs sequentially by solving problem (P2-m) based on SCA. Similarly, from step 10 to step 13, the positions of $N$ transmit MAs are optimized by solving problem (P3-n) successively. The algorithm proceeds by iteratively solving the three subproblems presented in Section III-A, until the increase of the channel capacity in \eqref{C} is below $\epsilon_2$, which is a predefined convergence threshold. 

\begin{algorithm}[!t]
	\caption{Alternating Optimization for Solving Problem (P1)}
	\label{alg2}
	\begin{algorithmic}[1]
		\STATE \emph{Input:} $\boldsymbol{\Sigma}$, $P$, $\sigma$, $M$, $N$, $L_r$, $L_t$, $\{\theta_r^q\}_{q=1}^{L_r}$, $\{\phi_r^q\}_{q=1}^{L_r}$, $\{\theta_t^p\}_{p=1}^{L_t}$, $\{\phi_t^p\}_{p=1}^{L_t}$, $\mathcal{C}_r$, $\mathcal{C}_t$, $D$, $\epsilon_1$, $\epsilon_2$.
        \STATE Initialize $\{\boldsymbol{r}_m\}_{m=1}^M$ and $\{\boldsymbol{t}_n\}_{n=1}^N$. 
        \WHILE{Increase of the channel capacity in \eqref{C} is above $\epsilon_2$}
        \STATE Obtain the optimal solution of $\boldsymbol{Q}$ to (P1) with given $\{\boldsymbol{r}_m\}_{m=1}^M$ and $\{\boldsymbol{t}_n\}_{n=1}^N$ according to \eqref{Qstar}.
        \FOR {$m=1\rightarrow M$}
        \STATE Obtain $\boldsymbol{B}_m$ via \eqref{Bm}.
        \STATE Given $\boldsymbol{Q}$, $\{\boldsymbol{r}_k, k\neq m\}_{k=1}^M$, and $\{\boldsymbol{t}_n\}_{n=1}^N$, solve problem (P2-m) to update $\boldsymbol{r}_m$.
        \ENDFOR
        \STATE Calculate $\boldsymbol{S}^{\star}$ via \eqref{Sstar}.
        \FOR {$n=1 \rightarrow N$}
        \STATE Obtain $\boldsymbol{D}_n$ via \eqref{Dn}.
        \STATE Given $\boldsymbol{Q}$, $\{\boldsymbol{t}_l, l\neq n\}_{l=1}^N$, and $\{\boldsymbol{r}_m\}_{m=1}^M$, solve problem (P3-n) to update $\boldsymbol{t}_n$.
        \ENDFOR
        \ENDWHILE

        \STATE \emph{Output:} $\tilde{\boldsymbol{t}}$, $\tilde{\boldsymbol{r}}$, $\boldsymbol{Q}$.
	\end{algorithmic}
\end{algorithm}

The convergence of the proposed Algorithm~\ref{alg2} is analyzed as follows. The alternating optimization of variables $\{\boldsymbol{t}_n\}_{n=1}^N$, $\{\boldsymbol{r}_m\}_{m=1}^M$, and $\boldsymbol{Q}$ guarantees that the algorithm yields non-decreasing objective value of (P1) during the iterations, which is also upper-bounded by a finite channel capacity. Moreover, since the objective function of (P1) is differentiable and all the variables $\{\boldsymbol{t}_n\}_{n=1}^N$, $\{\boldsymbol{r}_m\}_{m=1}^M$, and $\boldsymbol{Q}$ are not coupled in the constraints, any limit point of the iterations generated by
Algorithm~\ref{alg2} satisfies the Karush–Kuhn–Tucker (KKT) condition of (P1). By setting the convergence criterion of Algorithm~\ref{alg2} as that the objective function of (P1) cannot be further increased by optimizing any variable in $\{\boldsymbol{t}_n\}_{n=1}^N$, $\{\boldsymbol{r}_m\}_{m=1}^M$, and $\boldsymbol{Q}$, Algorithm~\ref{alg2} is guaranteed to converge to a (at least) locally optimal solution of (P1).

The computational complexity of Algorithm~\ref{alg2} is analyzed as follows. Specifically, in step 4 and step 9, the complexity to calculate $\boldsymbol{Q}^{\star}$ and $\boldsymbol{S}^{\star}$ based on water-filling is $\mathcal{O}(MN\min(M,N))$ \cite{zhang2020capacity}. Thus, the total complexity is $\mathcal{O}((MN L_r \gamma_r^1 + M^{2.5}\ln(1/\beta)\gamma_r^2 + MN L_t \gamma_t^1 + N^{2.5}\ln(1/\beta)\gamma_t^2 + MN\min(M,N))\gamma)$ with $\gamma$ denoting the maximum number of outer iterations for repeating steps 4–13 in Algorithm~\ref{alg2}.

\subsection{Initialization}

\begin{figure}[!t]
	\centering
	\includegraphics[width=90mm]{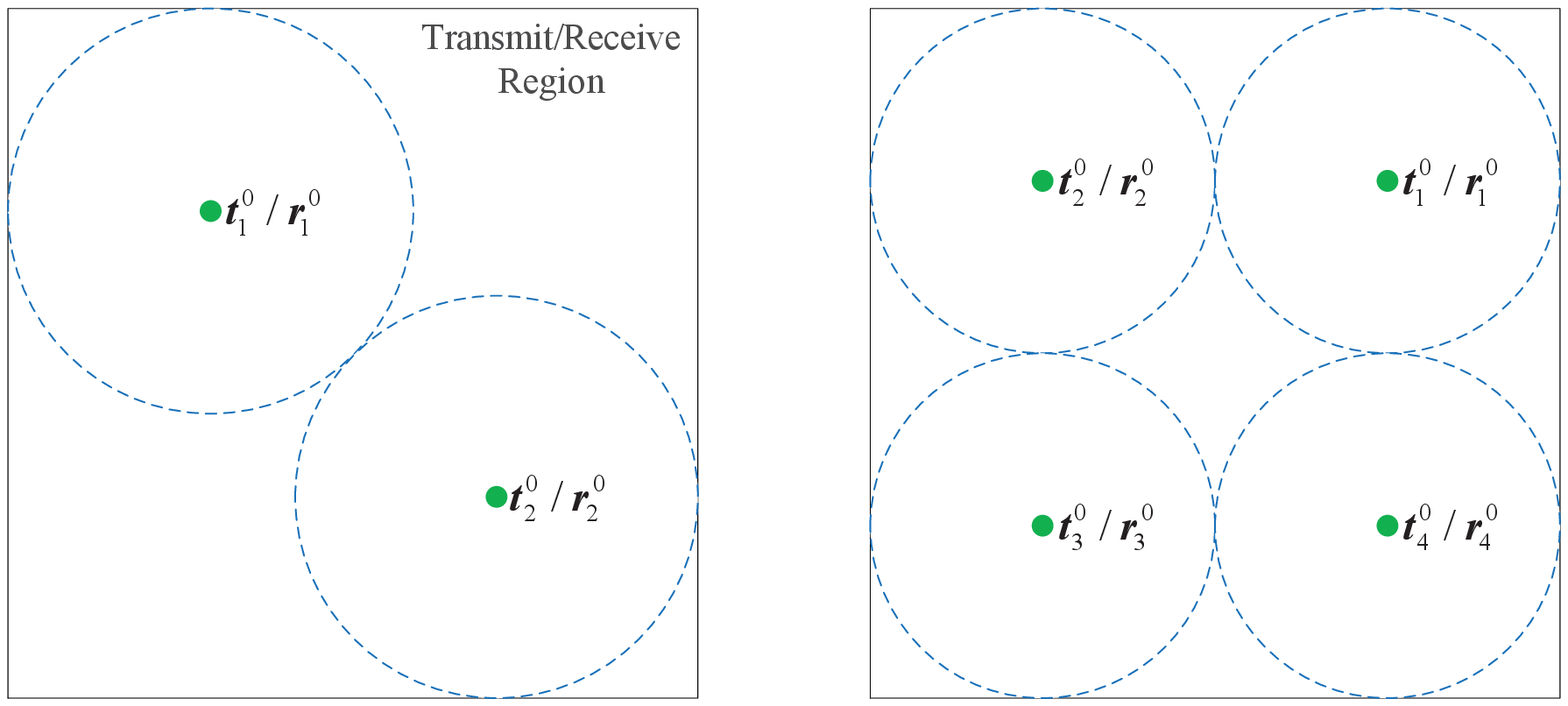}
	\caption{An example of transmit/receive MA positioning initialization based on circle packing for $N(M) = 2$ (left) and $N(M) = 4$ (right).}
	\label{FIG3}
\end{figure}

In this subsection, we propose a low-complexity initialization scheme for the transmit/receive MA positions design in Algorithm~\ref{alg2} based on the circle packing scheme \cite{packings2017}. Intuitively, all the transmit/receive MAs should be sufficiently separated for the following two reasons. On one hand, increasing the distance between MAs can reduce their coupling effect. One the other hand, during the SCA process, it is more likely for each MA to update the position with the largest capacity in its neighborhood satisfying the minimum distance constraints. Under such case, we can easily set $\boldsymbol{r}_{m,i+1}^{\star}$ in \eqref{rstar} as the optimal solution for problem (P4-m) instead of relaxing it into (P5-m). Therefore,
we propose that the initial positions of the MAs are obtained based on circle packing. As shown in Fig.~\ref{FIG3}, we assume that each MA occupies an initial circle region with equal radius. Then, we arrange $N$ (or $M$) circles with the maximum radius in $\mathcal{C}_t$ (or $\mathcal{C}_r$) such that no overlapping among them occurs. The centers of the arranged $N$ (or $M$) circles are set as the initial MA positions $\{\boldsymbol{t}_n^0\}_{n=1}^N$ (or $\{\boldsymbol{r}_m^0\}_{m=1}^M$).

\subsection{ Alternative Solution for Problem (P1) in the Low-SNR Regime}
In the previous subsections, we have proposed an alternating optimization algorithm that can handle the general MIMO channel capacity maximization problem (P1). In this subsection, we consider the special case with the asymptotically low-SNR regime, and derive the corresponding channel capacity in a more tractable form in terms of the positions of transmit and receive MAs, based on which a low-complexity alternative solution for (P1) is obtained.

The low-SNR regime may occur in practice due to the low transmission power or large propagation distance between the transmitter and the receiver. In such case, the optimal transmission strategy is to perform single-stream beamforming over the strongest eigenmode of the channel matrix $\boldsymbol{H}(\tilde{\boldsymbol{t}}, \tilde{\boldsymbol{r}})$ by allocating all the transmit power to the strongest eigenchannel \cite{goldsmith2005wireless}. Specifically, denoting $\boldsymbol{Q}=P\boldsymbol{u}_Q \boldsymbol{u}_Q^H$ with $\boldsymbol{u}_Q \in{\mathbb{C}^{N}}$ being the transmit beamforming vector, then the capacity in \eqref{C} can be rewritten as
\begin{align}\label{C_low}
	C_L(\tilde{\boldsymbol{t}}, \tilde{\boldsymbol{r}})&=\max_{\boldsymbol{u}_Q: \|\boldsymbol{u}_Q\|_2^2 = 1} \log_{2} \det\left(\boldsymbol{I}_M+\frac{P}{\sigma^2} \boldsymbol{H}(\tilde{\boldsymbol{t}}, \tilde{\boldsymbol{r}}) \boldsymbol{u}_Q \boldsymbol{u}_Q^H \boldsymbol{H}(\tilde{\boldsymbol{t}}, \tilde{\boldsymbol{r}})^H\right) \\
	&\overset{(d)}=\max_{\boldsymbol{u}_Q: \|\boldsymbol{u}_Q\|_2^2 = 1} \log_{2} \left(1+\frac{P}{\sigma^2}  \|\boldsymbol{H}(\tilde{\boldsymbol{t}}, \tilde{\boldsymbol{r}}) \boldsymbol{u}_Q\|_2^2\right), \notag
\end{align}
where the equality marked by $(d)$ can be derived because of $\det(\boldsymbol{I}_p+\boldsymbol{A}\boldsymbol{B}) = \det(\boldsymbol{I}_q+\boldsymbol{B}\boldsymbol{A})$ for $\boldsymbol{A} \in{\mathbb{C}^{p \times q}}$ and $\boldsymbol{B} \in{\mathbb{C}^{q \times p}}$. Based on \eqref{C_low}, the capacity maximization problem in the low-SNR
regime can be equivalently converted to maximizing  $\|\boldsymbol{H}(\tilde{\boldsymbol{t}}, \tilde{\boldsymbol{r}}) \boldsymbol{u}_Q\|_2^2$. Given $\boldsymbol{H}(\tilde{\boldsymbol{t}}, \tilde{\boldsymbol{r}})$, the optimal $\boldsymbol{u}_Q$ is the strongest right singular vector of $\boldsymbol{H}(\tilde{\boldsymbol{t}}, \tilde{\boldsymbol{r}})$. Besides, we denote $\boldsymbol{c} = \boldsymbol{\Sigma} \boldsymbol{G}(\tilde{\boldsymbol{t}}) \boldsymbol{u}_Q \in{\mathbb{C}^{L_r}}$. Then, we have $\|\boldsymbol{H}(\tilde{\boldsymbol{t}}, \tilde{\boldsymbol{r}}) \boldsymbol{u}_Q\|_2^2 = \sum_{m=1}^{M} |\boldsymbol{c}^H \boldsymbol{f}(\boldsymbol{r}_m)|^2$. Given $\boldsymbol{u}_Q$, $\{\boldsymbol{r}_k, k\neq m\}_{k=1}^M$, and $\{\boldsymbol{t}_n\}_{n=1}^N$, maximizing the channel capacity in \eqref{C_low} is equivalent to maximizing $|\boldsymbol{c}^H \boldsymbol{f}(\boldsymbol{r}_m)|^2$ with respect to each $\boldsymbol{r}_m$. Thus, the subproblem for optimizing $\boldsymbol{r}_m$ can be formulated similarly to problem (P2-m) by replacing $\boldsymbol{B}_m$ with $\boldsymbol{c} \boldsymbol{c}^H$, which has a lower computational complexity during the iterations. Moreover, given $\boldsymbol{u}_Q$, $\{\boldsymbol{t}_l, l\neq n\}_{l=1}^N$, and $\{\boldsymbol{r}_m\}_{m=1}^M$, the optimal $\boldsymbol{S}$ is given by $\boldsymbol{S}=P\boldsymbol{u}_S \boldsymbol{u}_S^H$ with $\boldsymbol{u}_S \in{\mathbb{C}^{M}}$ being the strongest right singular vector of $\boldsymbol{H}(\tilde{\boldsymbol{t}}, \tilde{\boldsymbol{r}})^H$. Defining $\boldsymbol{d} \triangleq \boldsymbol{\Sigma}^H \boldsymbol{F}(\tilde{\boldsymbol{r}}) \boldsymbol{u}_S \in{\mathbb{C}^{L_t}}$, the subproblem for optimizing $\boldsymbol{t}_n$ can be expressed similarly to problem (P3-n) by replacing $\boldsymbol{D}_n$ with $\boldsymbol{d} \boldsymbol{d}^H$.

Therefore, by employing a similar alternating optimization algorithm as Algorithm~\ref{alg2}, a locally optimal solution for the capacity maximization problem in the low-SNR regime can be obtained via iteratively optimizing the three sets of variables $\{\boldsymbol{t}_n\}_{n=1}^N$, $\{\boldsymbol{r}_m\}_{m=1}^M$, and $\boldsymbol{u}_Q$ with the other ones being fixed. The required complexity of the overall algorithm is given by $\mathcal{O}((L_r \gamma_r^1  + M^{2.5}\ln(1/\beta)\gamma_r^2 + L_t \gamma_t^1 + N^{2.5}\ln(1/\beta)\gamma_t^2 + MN\min(M,N))\gamma)$, which is generally lower than that of Algorithm~\ref{alg2} because $\boldsymbol{c}$ and $\boldsymbol{d}$ are fixed in the inner SCA iterations for solving problems (P2-m) and (P3-n).

\subsection{Alternative Solution for Problem (P1) with Single-Antenna Transmitter/Receiver}
So far, we have investigated (P1) for the general MIMO channel with $N\geq 1$ and $M\geq 1$, by considering parallel transmissions of multiple data streams in general. In this subsection, we study (P1) for the special cases with $N= 1$ or $M= 1$, where only one data stream can be transmitted. This
leads to more simplified expressions of the optimal transmit covariance matrix as well as the channel capacity, based on which we propose simpler alternating optimization algorithms for solving (P1) that require much lower complexity compared to Algorithm~\ref{alg2}.

First, we consider (P1) for the MISO case with $N > 1$ and $M= 1$, where the field response matrix of the receive region, $\boldsymbol{F}(\tilde{\boldsymbol{r}})$, can be rewritten as vector $\boldsymbol{f}(\boldsymbol{r}) \in{\mathbb{C}^{L_r}}$, where $\boldsymbol{r}$ represents the position of the receive MA. The effective channel row vector can be expressed as $\boldsymbol{h}(\tilde{\boldsymbol{t}}, \boldsymbol{r})^H = \boldsymbol{f}(\boldsymbol{r})^H \boldsymbol{\Sigma} \boldsymbol{G}(\tilde{\boldsymbol{t}}) \in{\mathbb{C}^{1\times{N}}}$. Note that in this case, the optimal transmit covariance matrix is given by the maximum ratio transmission (MRT), i.e., $\boldsymbol{Q}^{\star}=P \boldsymbol{h}(\tilde{\boldsymbol{t}}, \boldsymbol{r}) \boldsymbol{h}(\tilde{\boldsymbol{t}}, \boldsymbol{r})^H /\|\boldsymbol{h}(\tilde{\boldsymbol{t}}, \boldsymbol{r})\|_2^2$. Thus, the MISO channel capacity can be rewritten as
\begin{align}\label{C_MISO}
	C_{\rm{MISO}}(\tilde{\boldsymbol{t}}, \boldsymbol{r})&=\log_{2} \left(1+\frac{1}{\sigma^2} \boldsymbol{h}(\tilde{\boldsymbol{t}}, \boldsymbol{r})^H \boldsymbol{Q}^{\star} \boldsymbol{h}(\tilde{\boldsymbol{t}}, \boldsymbol{r})\right) \\
	&=\log_{2} \left(1+\frac{P}{\sigma^2} \|\boldsymbol{h}(\tilde{\boldsymbol{t}}, \boldsymbol{r}) \|_2^2\right). \notag
\end{align}
In this case, problem (P1) can be equivalently transformed into the following problem for maximizing the channel total power, $\|\boldsymbol{h}(\tilde{\boldsymbol{t}}, \boldsymbol{r}) \|_2^2$, via optimizing $\tilde{\boldsymbol{t}}$ and $\boldsymbol{r}$
\begin{align}\label{P1M}
	\textrm {(P1-MISO)}~~\max_{\tilde{\boldsymbol{t}}, \boldsymbol{r}} \quad & \|\boldsymbol{h}(\tilde{\boldsymbol{t}}, \boldsymbol{r}) \|_2^2 \\
	\text{s.t.} \quad & \eqref{P1b}, \eqref{P1c}, \eqref{P1d}. \notag
\end{align}

Given $\{\boldsymbol{t}_n\}_{n=1}^N$, the subproblem for optimizing $\boldsymbol{r}$ can be formulated similarly to problem (P2-m) by setting $\boldsymbol{B}_m \leftarrow \boldsymbol{\Sigma} \boldsymbol{G}(\tilde{\boldsymbol{t}}) \boldsymbol{G}(\tilde{\boldsymbol{t}})^H \boldsymbol{\Sigma}^H \in{\mathbb{C}^{L_r\times{L_r}}}$. Note that the non-convex minimum distance constraints \eqref{P1e} are removed since there is only a single MA in the receive region. Then, by dropping the corresponding non-convex constraints \eqref{P2c}, problem (P4-m) is a convex QP problem, which can be solved efficiently without relaxation into problem (P5-m). Moreover, we can obtain the closed-form solution for problem (P4-m) for some typical convex $\mathcal{C}_r$ based on $\boldsymbol{r}_{m,i+1}^{\star}=[x_{i+1}^\star, y_{i+1}^\star]^T$ in \eqref{rstar}. For example, if $\mathcal{C}_r$ is a rectangle region $\{[x, y]^T|x_L \leq x \leq x_H, y_L \leq y \leq y_H\}$, then the optimal solution for problem (P4-m) is given by \cite{fu2022uav}
\begin{equation}
	\boldsymbol{r}_m^{i+1} =[{\mathcal P_{\rm rec}} (x_{i+1}^\star, x_L, x_H), {\mathcal P_{\rm rec}} (y_{i+1}^\star, y_L, y_H)]^T,
\end{equation}
where ${\mathcal P_{\rm rec}} (x, x_L, x_H) = \min(\max(x, x_L), x_H)$ is the projection function associated with the rectangle region. Similarly, if $\mathcal{C}_r$ is a circle region $\{\boldsymbol{r}|\|\boldsymbol{r}-\boldsymbol{r}_0\|_2 \leq D_0\}$ centered at point $\boldsymbol{r}_0$ with radius of $D_0$, the optimal solution for problem (P4-m) can be written as
\begin{equation}
	\boldsymbol{r}_m^{i+1} ={\mathcal P_{\rm cir}} (\boldsymbol{r}_{m,i+1}^{\star}-\boldsymbol{r}_0, D_0)+\boldsymbol{r}_0,
\end{equation}
with the projector ${\mathcal P_{\rm cir}} (\boldsymbol{r}, D) = \min(D/\|\boldsymbol{r}\|_2, 1)\boldsymbol{r}$.

Moreover, we denote $\bar{\boldsymbol{d}} = \boldsymbol{\Sigma}^H \boldsymbol{f}(\boldsymbol{r})  \in{\mathbb{C}^{L_t}}$. Then, we have $\|\boldsymbol{h}(\tilde{\boldsymbol{t}}, \boldsymbol{r}) \|_2^2 = \| \bar{\boldsymbol{d}}^H \boldsymbol{G}(\tilde{\boldsymbol{t}}) \|_2^2 = \sum_{n=1}^{N} |\bar{\boldsymbol{d}}^H \boldsymbol{g}(\boldsymbol{t}_n)|^2$. Given $\boldsymbol{r}$ and $\{\boldsymbol{t}_l, l\neq n\}_{l=1}^N$, maximizing the channel total power in problem (P1-MISO) is equivalent to maximizing $|\bar{\boldsymbol{d}}^H \boldsymbol{g}(\boldsymbol{t}_n)|^2$ with respect to each $\boldsymbol{t}_n$. Thus, the subproblem for optimizing $\boldsymbol{t}_n$ can be formulated similarly to problem (P3-n) by replacing $\boldsymbol{D}_n$ with $\bar{\boldsymbol{d}} \bar{\boldsymbol{d}}^H$.

Therefore, a locally optimal solution to (P1-MISO) can be obtained via alternating optimization by iteratively optimizing one variable in $\boldsymbol{r} \cup \{\boldsymbol{t}_n\}_{n=1}^N$ with the other $N$ variables being fixed at each time. This algorithm requires a complexity of $\mathcal{O}((N L_r + L_r \gamma_r^1  +  L_t \gamma_t^1 + N^{2.5}\ln(1/\beta)\gamma_t^2)\gamma)$. It is worth noting that the complexity of this alternative algorithm is lower than that of Algorithm~\ref{alg2} with $M=1$, because the MISO channel capacity can be explicitly expressed as a function of $\tilde{\boldsymbol{t}}$ and $\boldsymbol{r}$ only, thus eliminating the need of iteratively solving $\boldsymbol{Q}$ in the alternating optimization.

Next, we consider the SIMO system with $M > 1$ and $N=1$, where the field response matrix of the transmit region $\boldsymbol{G}(\tilde{\boldsymbol{t}})$ can be rewritten as the vector $\boldsymbol{g}(\boldsymbol{t}) \in{\mathbb{C}^{L_t}}$ with $\boldsymbol{t}$ representing the position of the transmit MA. The SIMO channel is thus given by $\boldsymbol{h}(\boldsymbol{t}, \tilde{\boldsymbol{r}}) = \boldsymbol{F}(\tilde{\boldsymbol{r}})^H \boldsymbol{\Sigma} \boldsymbol{g}(\boldsymbol{t}) \in{\mathbb{C}^{M}}$. Note that with $N=1$, the optimal transmit covariance matrix can be easily obtained as $\boldsymbol{Q}^{\star}=P$. Therefore, the corresponding channel capacity is given by
\begin{align}\label{C_SIMO}
	C_{\rm{SIMO}}(\boldsymbol{t}, \tilde{\boldsymbol{r}})&=\log_{2} \det\left(\boldsymbol{I}_M+\frac{1}{\sigma^2} \boldsymbol{h}(\boldsymbol{t}, \tilde{\boldsymbol{r}}) \boldsymbol{Q} \boldsymbol{h}(\boldsymbol{t}, \tilde{\boldsymbol{r}})^H\right) \\
	&\overset{(e)}=\log_{2} \left(1+\frac{P}{\sigma^2} \|\boldsymbol{h}(\boldsymbol{t}, \tilde{\boldsymbol{r}}) \|_2^2\right),\notag
\end{align}
where the equality marked by $(e)$ holds because of $\det(\boldsymbol{I}_p+\boldsymbol{A}\boldsymbol{B}) = \det(\boldsymbol{I}_q+\boldsymbol{B}\boldsymbol{A})$ for $\boldsymbol{A} \in{\mathbb{C}^{p \times q}}$ and $\boldsymbol{B} \in{\mathbb{C}^{q \times p}}$. Notice that the above SIMO channel capacity is in a similar form as the MISO channel capacity $C_{\rm{MISO}}(\tilde{\boldsymbol{t}}, \boldsymbol{r})$, which can be maximized by maximizing the channel total power. Hence, the proposed alternating optimization algorithm for the MISO case can also be applied in the SIMO case with complexity $\mathcal{O}((L_r \gamma_r^1 + M^{2.5}\ln(1/\beta)\gamma_r^2 + M L_t + L_t \gamma_t^1)\gamma)$.

\section{Numerical Results}
In this section, we provide numerical results to evaluate the performance of our proposed algorithms for maximizing the MA-enabled MIMO channel capacity. In the simulation, we consider a MIMO system with $N=4$ transmit MAs and $M=4$ receive MAs. The transmit and receive regions are set as square areas with size $A\times A$. We consider the geometry channel model, where the numbers of transmit and receive paths are the same, i.e., $L_t = L_r$. The path response matrix is assumed to be diagonal with elements following i.i.d. CSCG distribution, i.e., $\boldsymbol{\Sigma}[p,p]\sim \mathcal{CN}(0, 1/L_r)$ for $p=1,2,\ldots,L_r$. The elevation and azimuth AoDs/AoAs are assumed to be i.i.d. variables with the uniform distribution over $[0, \pi]$. The minimum distance between MAs is set as $D=\lambda/2$. The average SNR is defined as $P/\sigma^2$. We set convergence thresholds for the relative increment of the objective value as $\epsilon_1=\epsilon_2=10^{-3}$ in Sections III-B and III-D. All the results are averaged over $4\times 10^4$ independent channel realizations. 

\begin{figure}[!t]
	\centering
	\includegraphics[width=70mm]{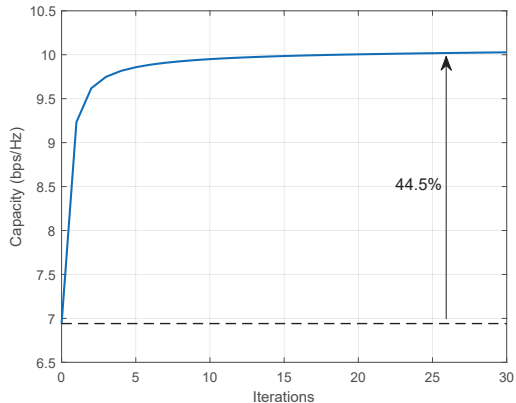}
	\caption{Convergence behavior of Algorithm~\ref{alg2}.}
	\label{FIG4}
\end{figure}

First, we set $L_t = L_r = 10$, $P/\sigma^2 = 5$ dB, $A=3\lambda$, and show in Fig.~\ref{FIG4} the convergence behavior of Algorithm~\ref{alg2}. It is observed that the capacity monotonically increases and converges to a maximum value around 10 bps/Hz after 20 iterations, which validates our analysis in Section III-D. Moreover, the converged capacity is increased by $44.5\%$ as compared to that at the initial point. 

Next, we compare the performance of Algorithm~\ref{alg2} with the following benchmark schemes:
\begin{enumerate}
	\item \textbf{FPA}: The transmitter and receiver are equipped with FPA-based uniform linear arrays (ULAs) with $N$ and $M$ antennas, respectively, spaced by $\lambda/2$.
	\item \textbf{AS}: The transmitter and receiver are equipped with FPA-based ULAs with $2N$ and $2M$ antennas, respectively, spaced by $\lambda/2$, where $N$ transmit antennas and $M$ receive antennas are selected via exhaustive search to maximize the channel capacity \cite{sanayei2004capacity}.
	\item \textbf{Strongest eigenchannel power maximization (SEPM)}: The alternative algorithm presented in Section III-F customized for the low-SNR regime.
	\item \textbf{Receive MA (RMA)}: The transmitter is equipped with an FPA-based ULA same as the FPA scheme, while the receiver employs $M$ MAs. Since $\{\boldsymbol{t}_n\}_{n=1}^N$ are constant with the FPA-based ULA, we only need to alternately optimize $\{\boldsymbol{r}_m\}_{m=1}^M  \cup \boldsymbol{Q}$ in Algorithm~\ref{alg2}.
	\item \textbf{Alternating position selection (APS)}: The transmit/receive area is quantized into discrete locations with equal-distance $D=\lambda/2$. The proposed Algorithm~\ref{alg2} is applied, where problems (P2-m) and (P3-n) are solved by exhaustive search over discrete locations for each transmit/receive MA. Note that it is difficult to perform exhaustive search over all possible transmit and receive MAs' positions due to the prohibitive computational complexity. For example, considering $N=M=4$ MAs in the $4\lambda\times4\lambda$ transmit/receive region, the number of total channel matrices for exhaustive search is $\tbinom{81}{4}^2=2.77\times 10^{12}$. As a result, we use Algorithm~\ref{alg2} to alternately select transmit and receive MA positions for lower complexity. 
\end{enumerate}

\begin{figure}[!t]
	\centering
	\subfigure[Capacity versus normalized region size.]{
		\begin{minipage}{.47\textwidth}
			\centering
			\includegraphics[scale=.5]{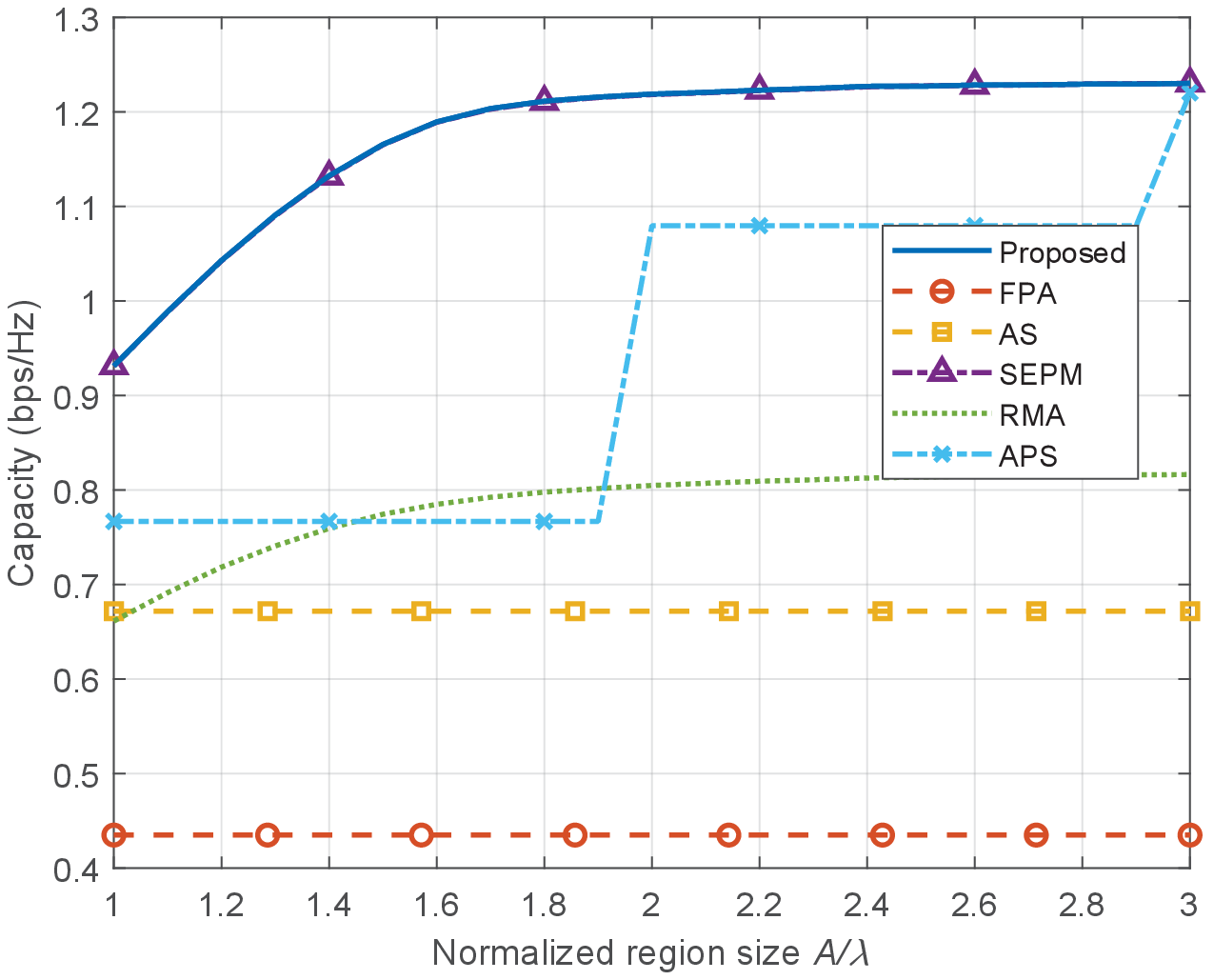}
		\end{minipage}
		\label{FIG5_1}
	}
	\subfigure[Strongest eigenchannel power versus normalized region size.]{
		\begin{minipage}{.47\textwidth}
			\centering
			\includegraphics[scale=.5]{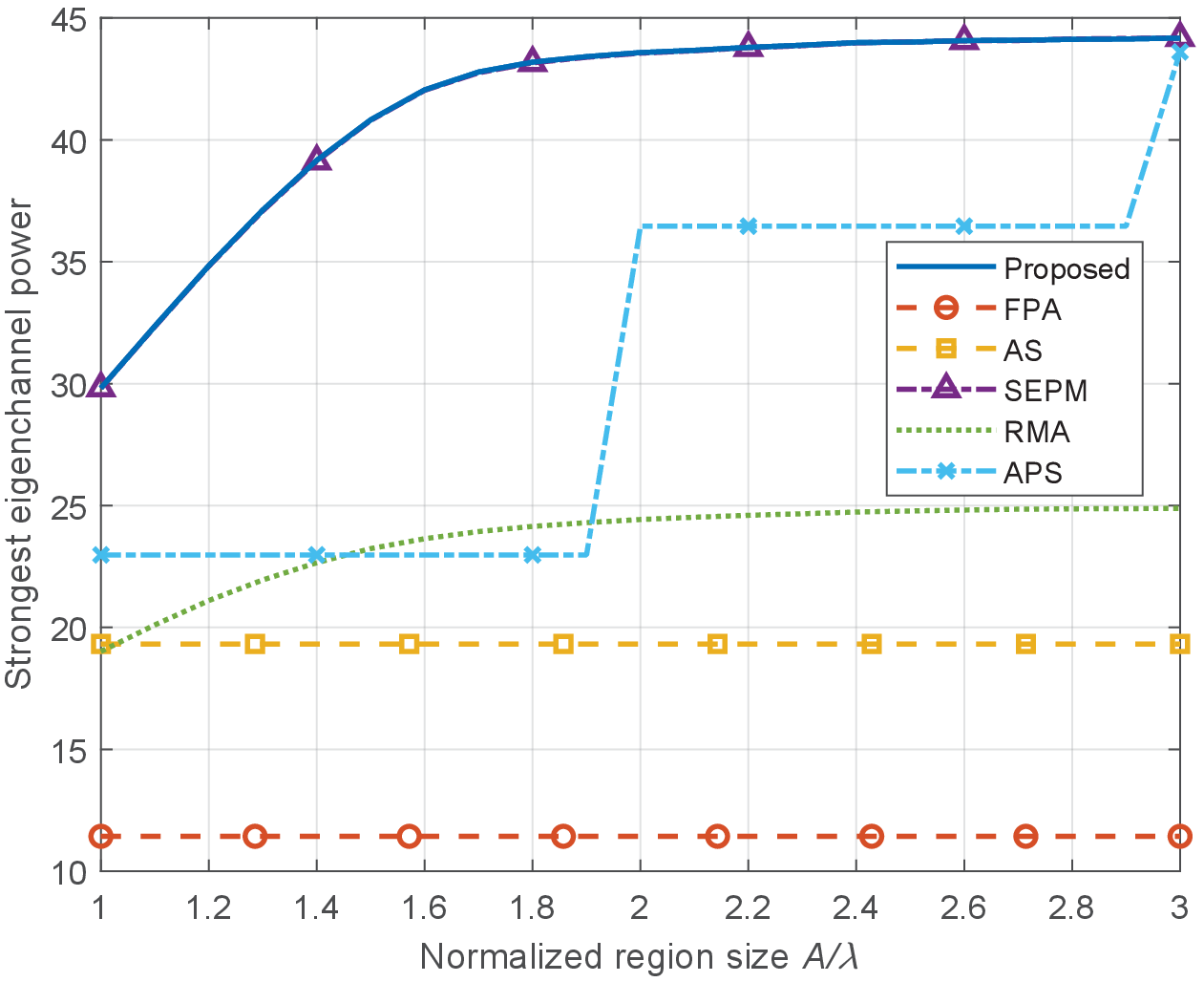}
		\end{minipage}
		\label{FIG5_2}
	}
	\caption{Performance of MA-enabled MIMO communication in the low-SNR regime.}
	\label{FIG5}
\end{figure}

\begin{figure}[!t]
	\centering
	\subfigure[Capacity versus normalized region size.]{
		\begin{minipage}{.31\textwidth}
			\centering
			\includegraphics[scale=.42]{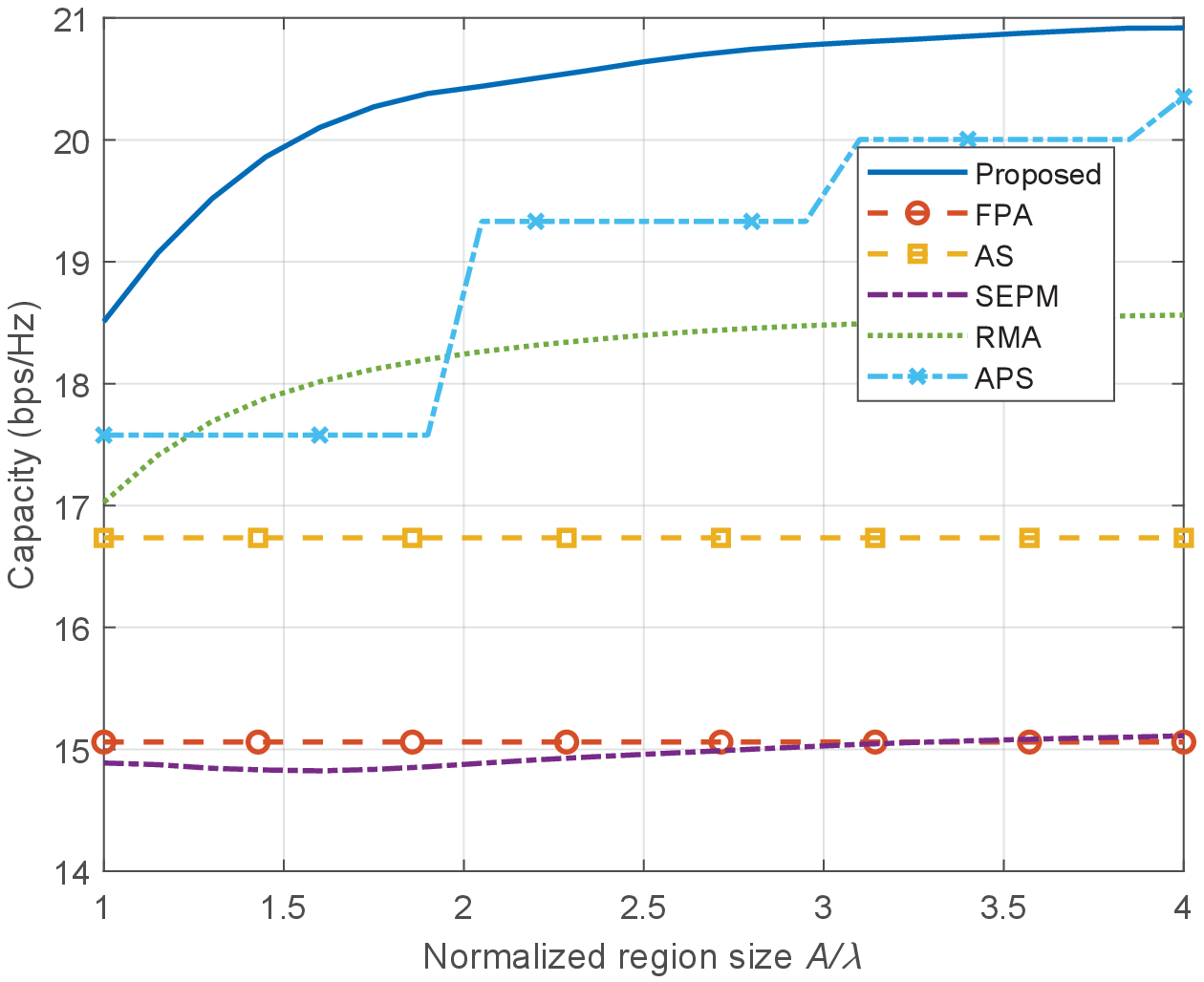}
		\end{minipage}
		\label{FIG6_1}
	}
	\subfigure[Channel total power versus normalized region size.]{
		\begin{minipage}{.31\textwidth}
			\centering
			\includegraphics[scale=.42]{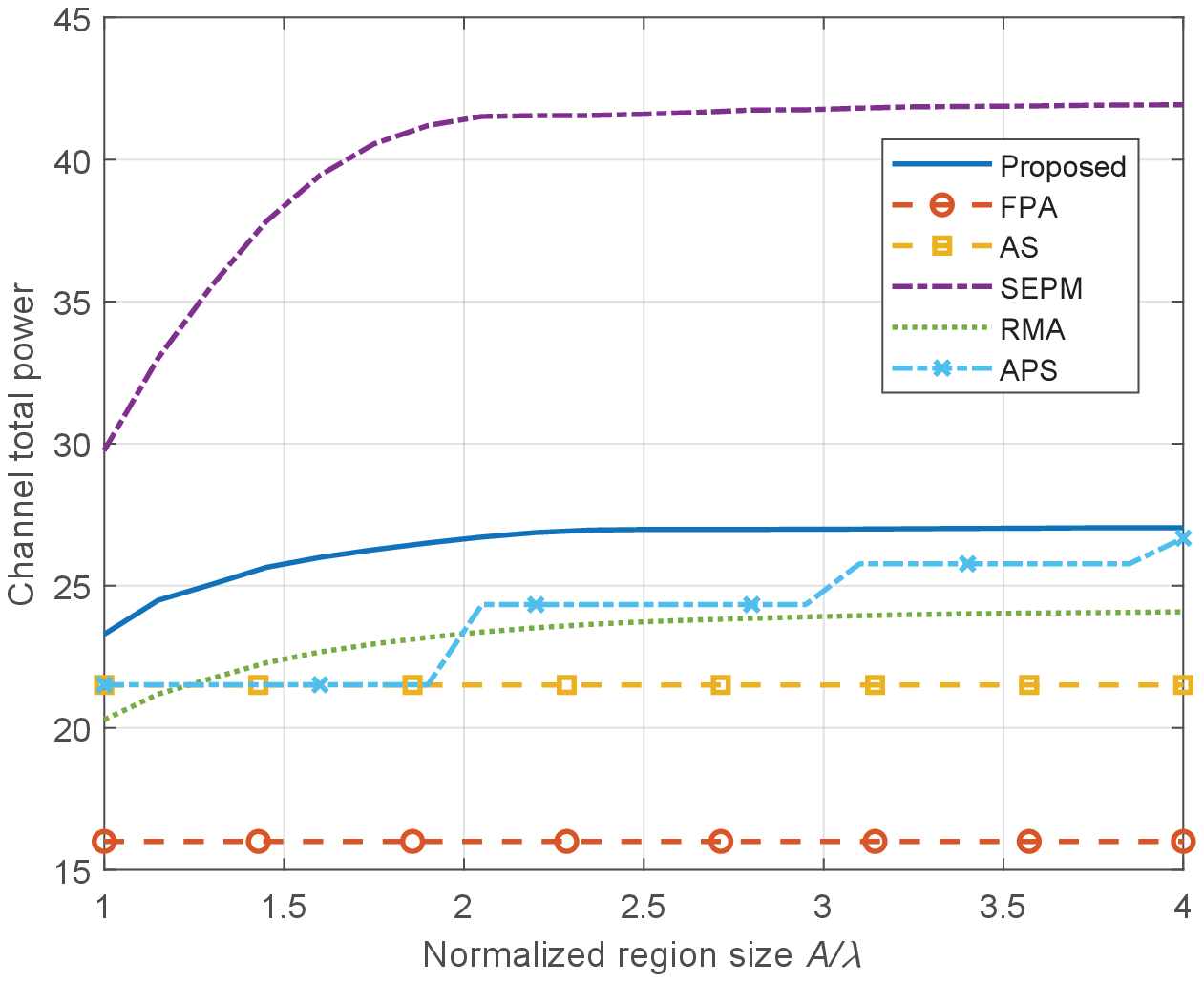}
		\end{minipage}
		\label{FIG6_2}
	}
	\subfigure[Channel condition number versus normalized region size.]{
		\begin{minipage}{.31\textwidth}
			\centering
			\includegraphics[scale=.42]{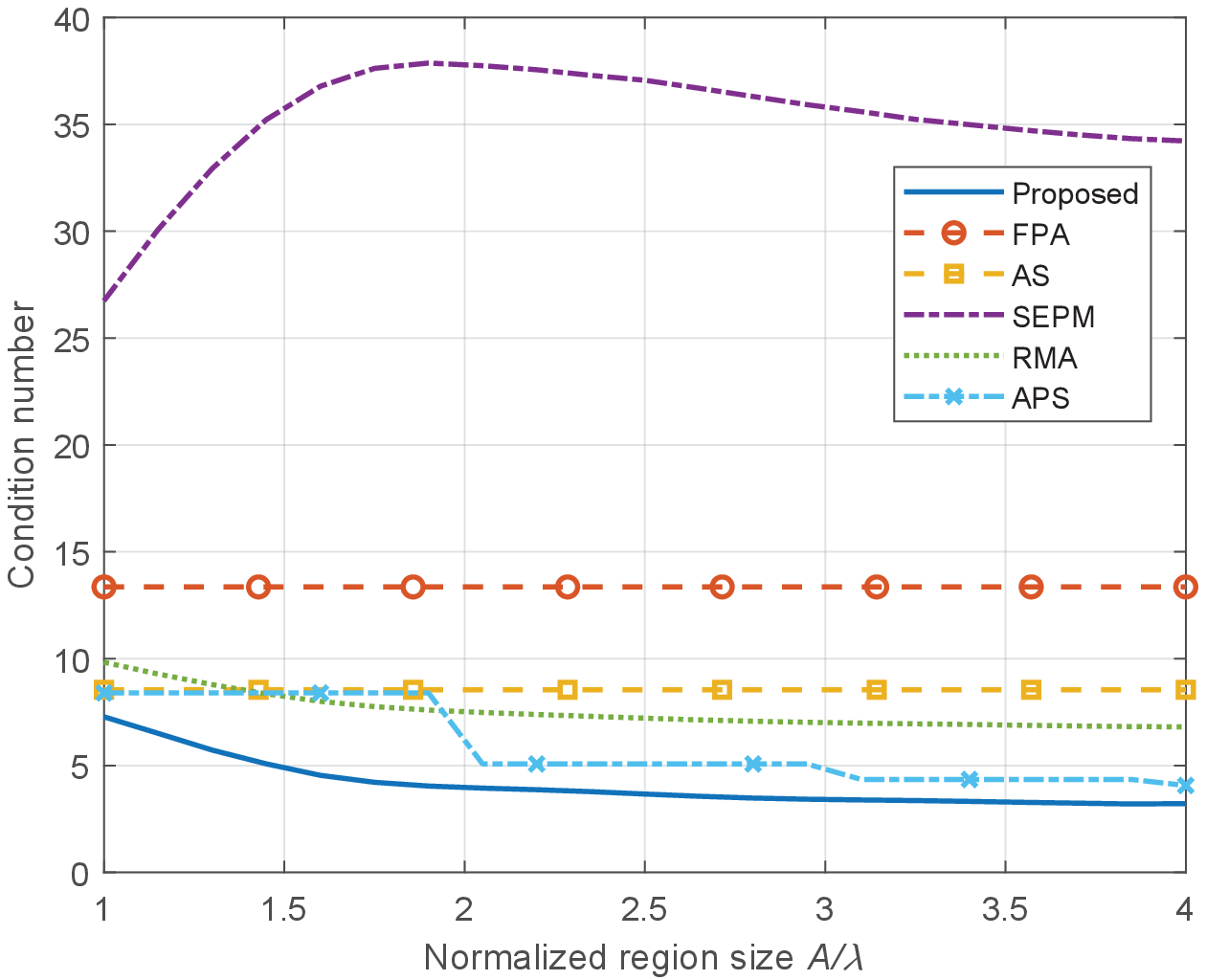}
		\end{minipage}
		\label{FIG6_3}
	}
	\caption{Performance of MA-enabled MIMO communication in the high-SNR regime.}
	\label{FIG6}
\end{figure}

We consider two SNR values of $-15$ dB and $15$ dB, which correspond to the low-SNR regime and high-SNR regime, respectively. We set $L_t = L_r = 10$. In Fig.~\ref{FIG5_1} and Fig.~\ref{FIG6_1}, we show the channel capacity versus the normalized region size $A/\lambda$ for the proposed MA-enabled MIMO system and the benchmark schemes. For both SNR regimes, it is observed that the proposed, RMA and APS schemes with MAs outperform FPA systems in terms of capacity, and the performance gain increases with the region size. Moreover, all schemes converge when the normalized region size is larger than $4$, which indicates that the maximum channel capacity of MA-enabled MIMO systems can be achieved with finite transmit and receive regions. It is also observed that our proposed algorithm achieves the best performance among all schemes in both low-SNR and high-SNR regimes for any region size. With $P/\sigma^2=15$ dB and $A/\lambda=3$, the proposed scheme achieves $38.1\%$, $24.3\%$, $38.3\%$, $12.5\%$, and $4.6\%$ performance improvements over the FPA, AS, SEPM, RMA, and APS schemes, respectively. Particularly, the proposed scheme with transmit and receive MAs outperforms the RMA scheme with only receive MAs, which shows the additional gain of jointly optimizing the transmit and receive MAs.

Moreover, in the low-SNR regime, it is observed from Fig.~\ref{FIG5_1} that the SEPM algorithm achieves almost the same performance as the proposed algorithm, which is consistent with our results in Section III-F. Thus, the SEPM scheme can serve as a low-complexity alternative to our proposed Algorithm~\ref{alg2} in the low-SNR regime. To draw more insight, we further depict in Fig.~\ref{FIG5_2} the average strongest eigenchannel power $\max(\textrm{diag}(\tilde{\boldsymbol{\Lambda}}))^2$ for each scheme versus the normalized region size. It is observed that the performance gap of $\max(\textrm{diag}(\tilde{\boldsymbol{\Lambda}}))^2$ between MA and FPA schemes increases with the normalized region size. However, in the high-SNR regime shown in Fig.~\ref{FIG6_1}, the SEPM scheme achieves the lowest channel capacity. The reason is as follows. Note that the channel total power and channel condition number $\zeta = \max(\textrm{diag}(\tilde{\boldsymbol{\Lambda}})) / \min(\textrm{diag}(\tilde{\boldsymbol{\Lambda}}))$ are two important parameters influencing the MIMO channel capacity in the high-SNR regime. Generally, the channel capacity increases with the channel total power and decreases with $\zeta$ \cite{goldsmith2005wireless}. From Fig.~\ref{FIG6_2} and Fig.~\ref{FIG6_3}, we observe that although the SEPM scheme can achieve the highest channel total power, it has the largest $\zeta$ as well. In other words, the SEPM scheme concentrates most of the channel power on the strongest eigenchannel but sacrifices the performance of the other eigenchannels, and thus the channel capacity is inferior as compared to other schemes in the high-SNR regime. It can be also observed that the channel total power of the proposed Algorithm~\ref{alg2} increases with the normalized region size and is higher than those of the FPA, AS, RMA, and APS schemes. Besides, the condition number of the proposed Algorithm~\ref{alg2} decreases with the normalized region size and is smaller than those of the FPA, AS, RMA, and APS schemes. The higher channel total power and lower condition number yield more balanced power distribution over different eigenchannels of the MIMO channel, thus leading to a capacity gain over other schemes. The above results indicate that our proposed Algorithm~\ref{alg2} is able to reshape the MIMO channel into a more favorable condition for capacity maximization.

\begin{figure}[!t]
	\centering
	\subfigure[$L_r=L_t=5$.]{
		\begin{minipage}{.47\textwidth}
			\centering
			\includegraphics[scale=.5]{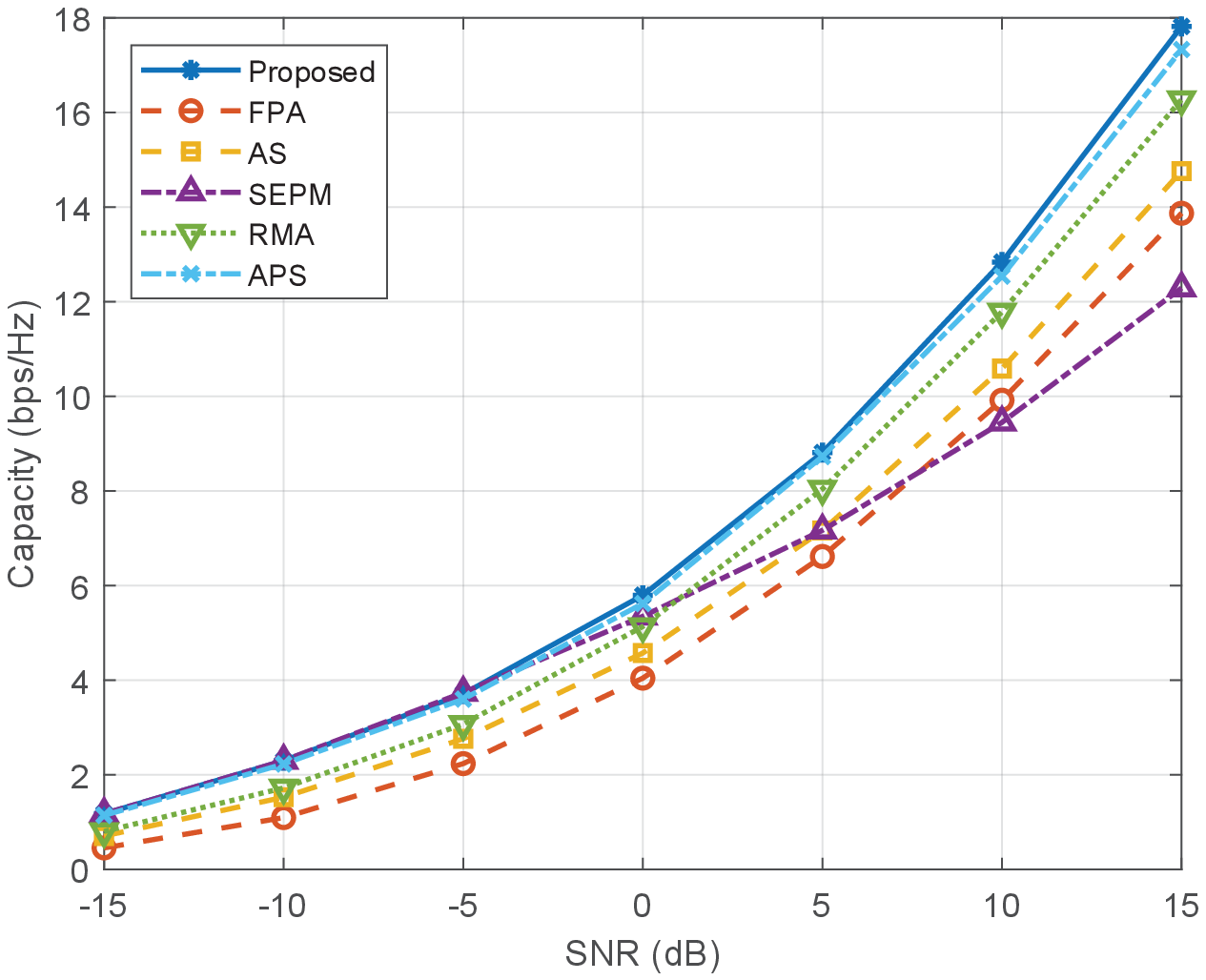}
		\end{minipage}
		\label{FIG7_1}
	}
	\subfigure[$L_r=L_t=15$.]{
		\begin{minipage}{.47\textwidth}
			\centering
			\includegraphics[scale=.5]{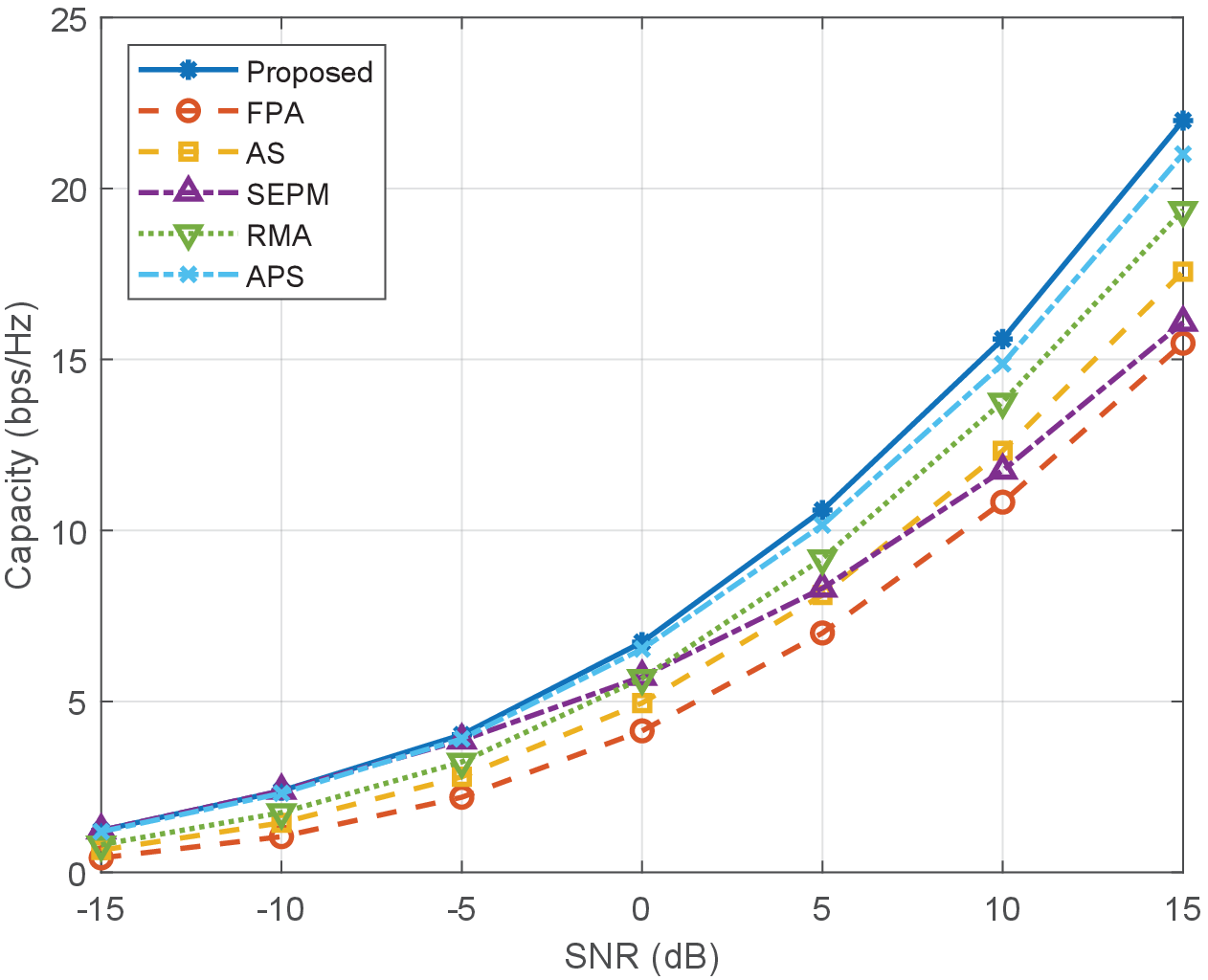}
		\end{minipage}
		\label{FIG7_2}
	}
	\caption{Capacity versus SNR for MA-enabled MIMO communication.}
	\label{FIG7}
\end{figure}

Finally, we consider a setup with $A=3\lambda$ and two different numbers of channel paths with $L_t = L_r = 5$ and
$L_t = L_r = 15$, respectively. In Fig.~\ref{FIG7}, we show the channel capacity of the proposed and benchmark schemes versus SNR. It is observed that with the same SNR, our proposed algorithm can achieve larger channel capacity as compared to the schemes with FPAs. For the case with $P/\sigma^2=15$ dB and $L_t=L_r=15$, the proposed scheme has $42.1\%$, $25.2\%$, $36.8\%$, $13.5\%$, and $4.7\%$ performance improvements over the FPA, AS, SEPM, RMA, and APS schemes, respectively. Moreover, it is observed from Fig.~\ref{FIG7} that the proposed and SEPM algorithms have almost the same performance when SNR is smaller than $-5$ dB, indicating that the SNR threshold of $-5$ dB guarantees that the proposed algorithm can be approximated by the SEPM scheme in the low-SNR regime. Moreover, for both two cases with different numbers of channel paths, the proposed, RMA and APS schemes yield a larger capacity compared to the FPA and AS schemes with FPAs, and the capacity gap increases with the number of channel paths. This is because the small-scale fading becomes stronger as the number of paths increases, and thus the channel capacity has more significant fluctuation in the transmit/receive region, which provides more DoFs on increasing the channel capacity for our proposed MA-enabled MIMO systems.

\section{Conclusions}
In this paper, we proposed a new MA-enabled MIMO system to increase the channel capacity by exploiting the antenna position optimization at the transmitter and receiver. We studied the capacity maximization problem for MA-enabled point-to-point MIMO communication, via the joint optimization of the transmit and receive MA positions as well as transmit covariance matrix. An alternating optimization algorithm was proposed to find a locally optimal solution by iteratively optimizing the position of each transmit/receive MA and the transmit covariance matrix with the other variables being fixed. Moreover, alternative algorithms of lower complexity were also proposed for the asymptotically low-SNR regime as well as MISO/SIMO channels. Numerical results showed that our proposed system and algorithm achieve significantly higher capacity over the conventional FPAs-based MIMO systems with/without AS in multi-path environment, especially when the transmit/receive region is sufficiently large. Moreover, it was revealed that by jointly optimizing transmit and receive MA positions, the MIMO channel power can be significantly improved while the condition number of the MIMO channel decreases, thus leading to more favorable MIMO channel realizations for enhancing the capacity.

\appendix
\subsection{Derivations of $\nabla \bar{g}(\boldsymbol{r}_m)$ and $\nabla^2 \bar{g}(\boldsymbol{r}_m)$}
For ease of exposition, we define
\begin{equation}\label{b}
  \boldsymbol{b} \triangleq \boldsymbol{B}_m \boldsymbol{f}(\boldsymbol{r}_m^i) \in{\mathbb{C}^{L_r}}.
\end{equation}
Further denote the $q$th entry of $\boldsymbol{b}$ as $b_q = |b_q|e^{j\angle{b_q}}$, with amplitude $|b_q|$ and phase $\angle{b_q}$. Thus, $\bar{g}(\boldsymbol{r}_m)$ can be written as
\begin{align}
    \bar{g}(\boldsymbol{r}_m) &= {\rm{Re}}\left\{ \boldsymbol{b}^H \boldsymbol{f}(\boldsymbol{r}_m)\right\} \\
    &={\rm{Re}}\left\{ \sum_{q=1}^{L_r} |b_q| e^{j\left(\frac{2\pi}{\lambda}\left(x_{r,m} \sin \theta_r^q \cos \phi_r^q + y_{r,m} \cos \theta_r^q\right)- \angle{b_q}\right)} \right\} \notag\\
    &=\sum_{q=1}^{L_r} |b_q| \cos\left(\kappa^q(\boldsymbol{r}_m)\right), \notag
\end{align}
with $\kappa^q(\boldsymbol{r}_m) \triangleq 2\pi\rho_r^q(\boldsymbol{r}_m)/\lambda- \angle{b_q}$. Recalling that $\boldsymbol{r}_m=[x_{r,m}, y_{r,m}]^T$, the gradient vector and Hessian matrix of $\bar{g}(\boldsymbol{r}_m)$ over $\boldsymbol{r}_m$ can be represented as $\nabla \bar{g}(\boldsymbol{r}_m) = \left[\frac{\partial \bar{g}(\boldsymbol{r}_m)}{\partial x_{r,m}}, \frac{\partial \bar{g}(\boldsymbol{r}_m)}{\partial y_{r,m}}\right]^T$ and $\nabla^2 \bar{g}(\boldsymbol{r}_m) = \begin{bmatrix}
   \frac{\partial \bar{g}(\boldsymbol{r}_m)}{\partial x_{r,m} \partial x_{r,m}} & \frac{\partial \bar{g}(\boldsymbol{r}_m)}{\partial x_{r,m} \partial y_{r,m}} \\
   \frac{\partial \bar{g}(\boldsymbol{r}_m)}{\partial y_{r,m} \partial x_{r,m}} & \frac{\partial \bar{g}(\boldsymbol{r}_m)}{\partial y_{r,m} \partial y_{r,m}}
  \end{bmatrix}$, respectively. Thus, we have
\begin{align}\label{gradg}
\frac{\partial \bar{g}(\boldsymbol{r}_m)}{\partial x_{m}} &= -\frac{2\pi}{\lambda}\sum_{q=1}^{L_r} |b_q| \sin \theta_r^q \cos \phi_r^q \sin\left(\kappa^q(\boldsymbol{r}_m)\right), \notag\\
\frac{\partial \bar{g}(\boldsymbol{r}_m)}{\partial y_{m}} &= -\frac{2\pi}{\lambda}\sum_{q=1}^{L_r} |b_q| \cos \theta_r^q \sin\left(\kappa^q(\boldsymbol{r}_m)\right), 
\end{align}
and
\begin{align}\label{grad2g}
\frac{\partial \bar{g}(\boldsymbol{r}_m)}{\partial x_{m} \partial x_{m}} &= -\frac{4\pi^2}{\lambda^2} \sum_{q=1}^{L_r} |b_q| \sin^2 \theta_r^q \cos^2 \phi_r^q \cos\left(\kappa^q(\boldsymbol{r}_m)\right), \notag\\
\frac{\partial \bar{g}(\boldsymbol{r}_m)}{\partial x_{m} \partial y_{m}} &= -\frac{4\pi^2}{\lambda^2} \sum_{q=1}^{L_r} |b_q| \sin \theta_r^q \cos \phi_r^q \cos \theta_r^q \cos\left(\kappa^q(\boldsymbol{r}_m)\right), \notag \\
\frac{\partial \bar{g}(\boldsymbol{r}_m)}{\partial y_{m} \partial x_{m}} &= -\frac{4\pi^2}{\lambda^2} \sum_{q=1}^{L_r} |b_q| \sin \theta_r^q \cos \phi_r^q \cos \theta_r^q \cos\left(\kappa^q(\boldsymbol{r}_m)\right), \notag \\
\frac{\partial \bar{g}(\boldsymbol{r}_m)}{\partial y_{m} \partial y_{m}} &= -\frac{4\pi^2}{\lambda^2} \sum_{q=1}^{L_r} |b_q| \cos^2 \theta_r^q \cos\left(\kappa^q(\boldsymbol{r}_m)\right).
\end{align}

\subsection{Construction of $\delta_m$}
Based on the definition of $\nabla^2 \bar{g}(\boldsymbol{r}_m)$ in Appendix A, we have
\begin{align}
	\|\nabla^2 \bar{g}(\boldsymbol{r}_m)\|_2^2 &\leq \|\nabla^2 \bar{g}(\boldsymbol{r}_m)\|_F^2 \\
	&=\left( \frac{\partial \bar{g}(\boldsymbol{r}_m)}{\partial x_{r,m} \partial x_{r,m}} \right)^2 +\left( \frac{\partial \bar{g}(\boldsymbol{r}_m)}{\partial x_{r,m} \partial y_{r,m}} \right)^2 +\left( \frac{\partial \bar{g}(\boldsymbol{r}_m)}{\partial y_{r,m} \partial x_{r,m}} \right)^2 +\left( \frac{\partial \bar{g}(\boldsymbol{r}_m)}{\partial y_{r,m} \partial y_{r,m}} \right)^2 \notag\\
	&\leq 4\left(\frac{4\pi^2}{\lambda^2} \sum_{q=1}^{L_r} |b_q|\right)^2. \notag
\end{align}
Since $\|\nabla^2 \bar{g}(\boldsymbol{r}_m)\|_2\boldsymbol{I}_2 \succeq \nabla^2 \bar{g}(\boldsymbol{r}_m)$, we can select $\delta_m$ as
\begin{equation}\label{delta}
  \delta_m = \frac{8\pi^2}{\lambda^2} \sum_{q=1}^{L_r} |b_q|,
\end{equation}
satisfying $\delta_m \geq \|\nabla^2 \bar{g}(\boldsymbol{r}_m)\|_2$, and thus we have $\delta_m\boldsymbol{I}_2 \succeq \nabla^2 \bar{g}(\boldsymbol{r}_m)$.

\bibliographystyle{IEEEtran}
\bibliography{IEEEabrv,IEEEexample}

\begin{thebibliography}{10}
\providecommand{\url}[1]{#1}
\csname url@samestyle\endcsname
\providecommand{\newblock}{\relax}
\providecommand{\bibinfo}[2]{#2}
\providecommand{\BIBentrySTDinterwordspacing}{\spaceskip=0pt\relax}
\providecommand{\BIBentryALTinterwordstretchfactor}{4}
\providecommand{\BIBentryALTinterwordspacing}{\spaceskip=\fontdimen2\font plus
\BIBentryALTinterwordstretchfactor\fontdimen3\font minus
  \fontdimen4\font\relax}
\providecommand{\BIBforeignlanguage}[2]{{%
\expandafter\ifx\csname l@#1\endcsname\relax
\typeout{** WARNING: IEEEtran.bst: No hyphenation pattern has been}%
\typeout{** loaded for the language `#1'. Using the pattern for}%
\typeout{** the default language instead.}%
\else
\language=\csname l@#1\endcsname
\fi
#2}}
\providecommand{\BIBdecl}{\relax}
\BIBdecl

\bibitem{jiang2021the}
W.~Jiang, B.~Han, M.~A. Habibi, and H.~D. Schotten, ``{The road towards 6G: A
  comprehensive survey},'' \emph{{IEEE} Open J. Commun. Soc.}, vol.~2, pp.
  334--366, Feb. 2021.

\bibitem{saad2020a}
W.~Saad, M.~Bennis, and M.~Chen, ``{A vision of 6G wireless systems:
  Applications, trends, technologies, and open research problems},''
  \emph{{IEEE} Netw.}, vol.~34, no.~3, pp. 134--142, May 2020.

\bibitem{chowdhury20206g}
M.~Z. Chowdhury, M.~Shahjalal, S.~Ahmed, and Y.~M. Jang, ``{6G wireless
  communication systems: Applications, requirements, technologies, challenges,
  and research directions},'' \emph{{IEEE} Open J. Commun. Soc.}, vol.~1, pp.
  957--975, 2020.

\bibitem{zhu2019milli}
L.~Zhu, J.~Zhang, Z.~Xiao, X.~Cao, D.~O. Wu, and X.~Xia, ``{Millimeter-wave
  NOMA with user grouping, power allocation and hybrid beamforming},''
  \emph{{IEEE} Trans. Wireless Commun.}, vol.~18, no.~11, pp. 5065--5079, Nov.
  2019.

\bibitem{lu2014an}
L.~Lu, G.~Y. Li, A.~L. Swindlehurst, A.~Ashikhmin, and R.~Zhang, ``{An overview
  of massive MIMO: Benefits and challenges},'' \emph{{IEEE} J. Sel. Topics
  Signal Process.}, vol.~8, no.~5, pp. 742--758, Oct. 2014.

\bibitem{sanayei2004antenna}
S.~Sanayei and A.~Nosratinia, ``{Antenna selection in MIMO systems},''
  \emph{{IEEE} Commun. Mag.}, vol.~42, no.~10, pp. 68--73, Oct. 2004.

\bibitem{gharavi2004fast}
M.~Gharavi-Alkhansari and A.~B. Gershman, ``{Fast antenna subset selection in
  MIMO systems},'' \emph{{IEEE} Trans. Signal Process.}, vol.~52, no.~2, pp.
  339--347, Feb. 2004.

\bibitem{wong2021fluid}
K.-K. Wong, A.~Shojaeifard, K.-F. Tong, and Y.~Zhang, ``{Fluid antenna
  systems},'' \emph{{IEEE} Trans. Wireless Commun.}, vol.~20, no.~3, pp.
  1950--1962, Mar. 2021.

\bibitem{wong2022fluid}
K.-K. Wong and K.-F. Tong, ``{Fluid antenna multiple access},'' \emph{{IEEE}
  Trans. Wireless Commun.}, vol.~21, no.~7, pp. 4801--4815, Jul. 2022.

\bibitem{zhang2022pattern}
Z.~Zhang and L.~Dai, ``{Pattern-division multiplexing for continuous-aperture
  MIMO},'' in \emph{Proc. IEEE ICC}, Seoul, Korea, Republic of, May 2022, pp.
  1--6.

\bibitem{huang2020holo}
C.~Huang, S.~Hu, G.~C. Alexandropoulos, A.~Zappone, C.~Yuen, R.~Zhang, M.~D.
  Renzo, and M.~Debbah, ``{Holographic MIMO surfaces for 6G wireless networks:
  Opportunities, challenges, and trends},'' \emph{{IEEE} Trans. Wireless
  Commun.}, vol.~27, no.~5, pp. 118--125, Oct. 2020.

\bibitem{pizzo2022fourier}
A.~Pizzo, L.~Sanguinetti, and T.~L. Marzetta, ``{Fourier plane-wave series
  expansion for holographic MIMO communications},'' \emph{{IEEE} Trans.
  Wireless Commun.}, vol.~21, no.~9, pp. 6890--6905, Sep. 2022.

\bibitem{decarli2021access}
N.~Decarli and D.~Dardari, ``{Communication modes with large intelligent
  surfaces in the near field},'' \emph{{IEEE} Access}, vol.~9, pp.
  165\,648--165\,666, Dec. 2021.

\bibitem{deng2021reconfi}
R.~Deng, B.~Di, H.~Zhang, Y.~Tan, and L.~Song, ``{Reconfigurable holographic
  surface: Holographic beamforming for metasurface-aided wireless
  communications},'' \emph{{IEEE} Trans. Veh. Technol.}, vol.~70, no.~6, pp.
  6255--6259, Jun. 2021.

\bibitem{wei2022multi}
L.~Wei, C.~Huang, G.~Alexandropoulus, W.~E. Sha, Z.~Zhang, M.~Debbah, and
  C.~Yuen, ``{Multi-user holographic MIMO surfaces: Channel modeling and
  spectral efficiency analysis},'' \emph{{IEEE} J. Sel. Top. Signal Process.},
  vol.~16, no.~5, pp. 1112--1124, Aug. 2022.

\bibitem{ismial1991null}
T.~Ismail and M.~Dawoud, ``{Null steering in phased arrays by controlling the
  elements positions},'' \emph{{IEEE} Trans. Antennas Propagat.}, vol.~39,
  no.~11, pp. 1561--1566, Nov. 1991.

\bibitem{basbug2017design}
S.~Basbug, ``{Design and synthesis of antenna array with movable elements along
  semicircular paths},'' \emph{{IEEE} Antennas Wireless Propag. Lett.},
  vol.~16, pp. 3059--3062, Oct. 2017.

\bibitem{zhuravlev2015experi}
A.~Zhuravlev, V.~Razevig, S.~Ivashov, A.~Bugaev, and M.~Chizh, ``{Experimental
  simulation of multi-static radar with a pair of separated movable
  antennas},'' in \emph{Proc. {IEEE} COMCAS}, Tel Aviv, Israel, Nov. 2015, pp.
  1--5.

\bibitem{goldsmith2005wireless}
A.~Goldsmith, \emph{{Wireless Communications}}.\hskip 1em plus 0.5em minus
  0.4em\relax Cambridge, U.K.: Cambridge Univ. Press, 2005.

\bibitem{petersen2006matrix}
K.~B. Petersen and M.~S. Pedersen, \emph{{The Matrix Cookbook}}.\hskip 1em plus
  0.5em minus 0.4em\relax Kgs. Lyngby, Denmark: Tech. Univ. Denmark, 2006.

\bibitem{wu2018joint}
Q.~Wu, Y.~Zeng, and R.~Zhang, ``{Joint trajectory and communication design for
  multi-UAV enabled wireless networks},'' \emph{{IEEE} Trans. Wireless
  Commun.}, vol.~17, no.~3, pp. 2109--2121, Mar. 2018.

\bibitem{magnus1995matrix}
J.~R. Magnus and H.~Neudecker, \emph{{Matrix Differential Calculus With
  Applications in Statistics and Econometrics}}.\hskip 1em plus 0.5em minus
  0.4em\relax Hoboken, NJ, USA: Wiley, 1995.

\bibitem{turlach2007quad}
B.~A. Turlach and A.~Weingessel, ``{quadprog: Functions to solve quadratic
  programming problems},'' \emph{CRAN-Package quadprog}, 2007.

\bibitem{ben2001lecture}
A.~Ben-Tal and A.~Nemirovski, \emph{{Lectures on Modern Convex Optimization:
  Analysis, Algorithms and Engineering Applications}}.\hskip 1em plus 0.5em
  minus 0.4em\relax Philadelphia, PA, USA: SIAM, 2001.

\bibitem{zhang2020capacity}
S.~Zhang and R.~Zhang, ``{Capacity characterization for intelligent reflecting
  surface aided MIMO communication},'' \emph{{IEEE} J. Sel. Areas Commun.},
  vol.~38, no.~8, pp. 1823--1838, Aug. 2020.

\bibitem{packings2017}
\emph{Packings of Equal Circles in Fixed-Sized Containers with Maximum Packing
  Density}, Accessed: Apr. 5, 2017. [Online]. Available:
  http://www.packomania.com.

\bibitem{fu2022uav}
M.~Fu, Y.~Zhou, Y.~Shi, W.~Chen, and R.~Zhang, ``{UAV aided over-the-air
  computation},'' \emph{{IEEE} Trans. Wireless Commun.}, vol.~21, no.~7, pp.
  4909--4924, Jul. 2022.

\bibitem{sanayei2004capacity}
S.~Sanayei and A.~Nosratinia, ``{Capacity maximizing algorithms for joint
  transmit–receive antenna selection},'' in \emph{Proc. IEEE ACSSC}, Pacific
  Grove, CA, USA, Nov. 2004, pp. 1773--1776.

\end{thebibliography}
\end{document}